\documentclass[a4paper,11pt]{article} 
\pdfoutput=1
\usepackage{jheppub} 
\usepackage{multirow}

\usepackage{color}
\usepackage{graphicx}
\usepackage{array}
\usepackage{xcolor}
\usepackage[normalem]{ulem}   
\usepackage{subcaption}

\newcommand{\xiu}[1]{\textcolor{blue}{#1}}

 \usepackage{bm}

\newcommand{\be}{\begin{equation}}
\newcommand{\ee}{\end{equation}}
\newcommand{\bea}{\begin{eqnarray}}
\newcommand{\eea}{\end{eqnarray}}

\title{\boldmath Systematic Study of Coupled-Channel Dynamics in Doubly Heavy Hadronic Molecules}

\author[a]{Yu-Shan Ren,}
\author[b,1]{Guang-Juan Wang,\note{Corresponding author.}}
\author[a,1]{Zhi Yang,}
\author[c,d]{and Jia-Jun Wu}

\affiliation[a]{School of Physics, University of Electronic Science and
Technology of China, \\Chengdu 610054, China}
\affiliation[b]{Advanced Science Research Center, Japan Atomic Energy
Agency, \\Tokai, Ibaraki, 319-1195, Japan}
\affiliation[c]{School of Physical Sciences, University of Chinese Academy of Sciences (UCAS), \\Beijing 100049, China}
\affiliation[d]{Southern Center for Nuclear-Science Theory (SCNT), Institute of Modern Physics, Chinese Academy of Sciences, Huizhou 516000, China}

\emailAdd{wgj@pku.edu.cn}
\emailAdd{zhiyang@uestc.edu.cn}

\abstract{
Heavy Quark Spin Symmetry (HQSS) is widely use to predict heavy molecules by extending the effective interactions fitted from low-lying states to heavier sectors.  
In this work, we systematically investigate the reliability of this approach for higher double heavy tetraquarks by comparing a single-channel effective interaction (Scheme I) with an explicit coupled-channel dynamics framework (Scheme II).  
The interactions are obtained within one-boson-exchange potential model and fixed by fitting the $T_{cc}^+$ lineshape. 
Utilizing the complex scaling method and $T$-matrix pole analysis, we extract the possible poles in the $S$-wave $D^{(*)}D^{(*)}$, $\bar{B}^{(*)}\bar{B}^{(*)}$ and $D^{(*)}\bar{B}^{(*)}$ systems with $J^{P}=1^+$. 
We find that both schemes provide consistent descriptions of the lowest-lying state. 
This confirms isoscalar-dominated $T_{cc}$ as a predominant $DD^*$ molecule (binding energy $\sim$ 381 keV), and predicts an isoscalar deeply bound $T_{bb}$ state ($40-60$ MeV) and an isovector $T^\prime_{bb}$ resonance in the bottom sector, together with a virtual $T_{bc}$ state. 
In contrast, significant differences emerge for higher-lying states. 
The inclusion of explicit coupled-channel dynamics modifies the effective interaction and reshapes the pole structure. 
The states predicted as bound or resonant in the single-channel framework can be shifted far from the physical region or disappear.
These results indicate that while single-channel descriptions are adequate for near-threshold states,  an explicit treatment of coupled-channel dynamics is required for reliable predictions of excited doubly heavy tetraquarks.}

\begin{document} 
\maketitle
\flushbottom

\section{Introduction}
Conventional mesons and baryons are composed of quark–antiquark pairs or three quarks, respectively. 
However, a growing number of near-threshold exotic hadronic states discovered in recent years have continuously challenged this conventional picture. 
These states provide a unique testing ground for nonperturbative quantum chromodynamics (QCD) and offer crucial insights into the nature of the strong interaction in the low-energy regime, where confinement and dynamical binding mechanisms dominate.
Among them, the doubly charmed tetraquark $T_{cc}^{+}$, observed by the LHCb collaboration~\cite{LHCb:2021vvq,LHCb:2021auc}, stands out as an clean probe of the near-threshold dynamics.
Observed in the \(D^{0} D^{0} \pi^{+}\) invariant mass spectrum, this state is located just below the $D^0D^{*+}$ threshold with a mass difference of $M_{T_{cc}^{+}} - M_{D^{*+}D^0} = -360\pm40^{+4}_{-0}$ keV and an extremely narrow width $48\pm2^{+0}_{-14}$ keV.
Its proximity to threshold implies a strong sensitivity to coupled-channel effects, making it an ideal system to provide indispensable insights into the dynamical origin of hadronic molecules. 
Further details can be found in the recent extensive reviews~\cite{Esposito:2016noz,Chen:2022asf,Chen:2016qju,Guo:2017jvc,Meng:2022ozq,Bicudo:2022cqi}, and references therein.

The $T_{cc}^{+}$ is widely interpreted as an isoscalar $I=0$ bound state just below the $D^0D^{*+}$ threshold with spin-parity quantum numbers $J^P = 1^+$~\cite{Li:2012ss,Liu:2019stu,Dong:2021bvy,Dai:2021vgf,Ortega:2022efc,Asanuma:2023atv}. 
Following its discovery, the existence of its possible partner states in the \(D^* D^*\) system~\cite{Albaladejo:2021vln, Du:2021zzh,Lu:2025zae, Montesinos:2025mfx}, as well as in the doubly bottom ($T_{bb}$)~\cite{Albaladejo:2021vln,Du:2021zzh,Lu:2025zae, Montesinos:2025mfx,Bicudo:2015kna,Bicudo:2016ooe,Aoki:2023nzp} and mixed-flavor ($T_{bc}$) sectors
~\cite{Lee:2009rt,Chen:2013aba,Karliner:2017qjm,Sakai:2017avl,Carames:2018tpe,Richard:2022fdc,Liu:2025fhl,Song:2023izj,Lu:2020rog,Park:2018wjk,Eichten:2017ffp,Ebert:2007rn}, 
have been extensively discussed based on heavy quark spin symmetry (HQSS) and heavy quark flavor symmetry (HQFS). 
In many phenomenological approaches, the effective interaction is first constrained by fitting the lowest-lying state (e.g., $DD^*$ interactions for the $T_{cc}^+$) and then extrapolated to higher-mass systems using HQSS and HQFS, such as the $D^*D^*$ or the doubly bottom $T_{bb}$ and mixed $T_{bc}$ systems.
In such a single-channel framework, the effects of nearby channels are not treated explicitly but are effectively encoded into short-range interactions and regulator dependence.
While such an approach offers reasonable leading-order estimations, its applicability to higher-lying states may be questionable. 
In the heavy-hadron sector, multiple thresholds often lie within a narrow energy region close to the states of interest.
The coupled-channel dynamics can become an essential part of the physical degrees of freedom rather than a subleading correction \cite{Nieves:2012tt}. 
In such cases, absorbing these effects into short-range interactions may become inadequate when predicting higher-lying states. 
In particular, when the mass splittings between spin-multiplet thresholds become comparable to the binding energies, the predicted pole structure may exhibit strong model dependence, limiting the reliability of single-channel extrapolations to higher-lying states.

The limitations of single-channel approximations and the necessity of explicit coupled-channel dynamics have been recognized in hadron physics. 
A well-known example is the $\Lambda(1405)$, which arises from the strong coupling between $\bar{K}N$ and $\pi\Sigma$ channels, leading to a dynamically generated two-pole structure~\cite{Oset:1997it, Oller:2000fj, Jido:2003cb, Hyodo:2011ur}.
In this case, any attempt to describe it with a single-channel effective potential fails to produce the correct analytic structure. 
Similarly, in the hidden-charm sector, the extrapolation from the $X(3872)$ to its spin partner $X_2(D^*\bar D^*)$ requires an explicit coupled-channel treatment.
Such effects have been shown to significantly modify both the mass and decay width~\cite{Hidalgo-Duque:2012rqv, Guo:2013sya, Baru:2016iwj}.
Analogous effects are found in the hidden-bottom sector for the $Z_b(10610)$ and $Z'_b(10650)$ states~\cite{Baru:2017gwo,Mehen:2011yh,Coito:2016ads}. 
Because these twin resonances reside precisely at the $B\bar{B}^*$ and $\bar{B}^*\bar{B}^*$ thresholds respectively, an explicit coupled-channel mechanism is mandatory to simultaneously generate both states. 

Driven by these insights, doubly heavy tetraquark states provide an ideal platform to study the interplay between explicit coupled-channel dynamics and extrapolations based on HQSS, with the three representative systems, $T_{cc}$, $T_{bb}$, and $T_{bc}$, exhibiting distinct characteristics due to their different heavy quark compositions and offering a unique comparative framework to test strong interaction theories. 
The $T^+_{cc}$ state, as the only experimentally confirmed doubly heavy tetraquark to date, serves as a critical experimental input to constrain the hadronic interaction, and in the charm sector, the mass splitting between vector and pseudoscalar mesons is $\Delta_c = M_{D^*} - M_D \approx 142$ MeV, which defines the baseline for coupled-channel effects in doubly heavy systems. 
In the bottom sector, the $T_{bb}$ system exhibits exceptionally good HQSS due to the large bottom-quark mass; this property not only suppresses the kinetic energy and favors the formation of bound states~\cite{Ren:2023pip,Bicudo:2015kna}, but also leads to a much smaller mass splitting between the $\bar{B}\bar{B}^*$ and $\bar{B}^*\bar{B}^*$ thresholds ($\sim 46$ MeV) compared to the charm sector, such proximity significantly enhances coupled-channel effects and can induce substantial shifts in the pole positions. 
Numerous phenomenological works have been devoted to these states over the past two decades~\cite{Silvestre-Brac:1993zem,Semay:1994ht,Zhang:2007mu,Du:2012wp,Francis:2016hui,Karliner:2017qjm,Eichten:2017ffp,Junnarkar:2018twb,Leskovec:2019ioa,Cheng:2020wxa,Mohanta:2020eed,Meng:2021jnw,Chen:2021vhg,Ling:2021bir,Ren:2021dsi,Feijoo:2021ppq,Yan:2021wdl,Chen:2021tnn,Kim:2022mpa,Hudspith:2023loy,Zhang:2025vew}, with multiple studies based on single-channel $\bar{B}\bar{B}^*$ potentials predicting a deeply bound $J^P = 1^+$ state~\cite{Li:2012ss,Sakai:2025djx,Dong:2021bvy,Dong:2021juy}, and lattice QCD calculations further supporting a deeply bound $I(J^P) = 0(1^+)$ state with binding energies ranging from tens to a few hundred MeV~\cite{Bicudo:2015kna,Bicudo:2016ooe,Aoki:2023nzp}. 
In contrast, the mixed charm-bottom sector ($T_{bc}$) possesses an even richer structure than the pure charm or pure bottom systems, involving multiple nearby thresholds including $D\bar{B}^*$, $D^*\bar{B}$, and $D^*\bar{B}^*$, which leads to far more intricate coupled-channel dynamics and consequently significant discrepancies in theoretical predictions: some works predict shallow bound states driven by inter-channel couplings~\cite{Lee:2009rt,Chen:2013aba,Karliner:2017qjm,Sakai:2017avl,Carames:2018tpe,Richard:2022fdc,Liu:2025fhl}, while others find repulsive interactions in certain light-quark configurations that hinder the formation of bound states~\cite{Song:2023izj,Lu:2020rog,Park:2018wjk,Eichten:2017ffp,Ebert:2007rn}, and recent lattice QCD studies also present conflicting results, with some reporting a shallow bound state below threshold in the $D\bar{B}^{(*)}$ system~\cite{Alexandrou:2023cqg,Padmanath:2023rdu} whereas a dedicated search found no significant signal for a compact $bc\bar{u}\bar{d}$ tetraquark~\cite{Meinel:2022lzo}. 
Comparing the evolution of pole structures from the $D$ to $B$ sectors provides a clear way to quantify how coupled-channel dynamics reshape the hadron spectrum, and the distinct characteristics and current research status of these three systems highlight the necessity of a unified framework to consistently quantify coupled-channel effects across all three families, which is essential for making reliable predictions of their spectra and observable properties.

In this work, we systematically investigate the reliability of HQSS extrapolations for doubly heavy tetraquarks by comparing two distinct schemes.
In Scheme I, the interactions is treated with an artificially decoupled single-channel framework, where the $PP^*$ interaction ($P=D,B$) is retained and the $PP^*-P^*P^*$ coupling is effectively absorbed into short-range parameters \footnote{Here, ``single-channel" refers to the $PP^*$ channel space.
The coupling between different charged components, such as $D^+D^{*0}$ and $D^0D^{*+}$, which leads to isospin mixing due to the mass splitting, is still taken into account}. 
In contrast, Scheme II explicitly incorporates the coupled-channel dynamics between $PP^*$ and $P^*P^*$ channels and constrains the interaction through a refit to the $T_{cc}^+$ lineshape. 
The interactions between heavy mesons are constructed within the one-boson-exchange (OBE) potential model. 
It incorporates the exchange of light mesons, such as the pseudoscalar $\pi$ and the vector $\rho$ and $\omega$, to mediate inter-meson forces.
Rather than relying on arbitrary parameter tuning, in this study, we address refining the OBE coupling constants through a fit to the observed $T_{cc}^+$ lineshape in both schemes. 
According to HQFS, the leading-order effective potentials are largely independent of the heavy quark mass. 

In the previous work, we considered the single-channel description, which relies on a strict separation of scales, where only short-distance dynamics are safely integrated out and absorbed into contact terms. 
However, absorbing coupled-channel effects into static short-range parameters implicitly assumes that the intermediate-channel momentum scale $p \sim \sqrt{2\mu \Delta M}$ (where $\mu$ is the reduced mass and $\Delta M$ the threshold mass splitting of the coupled channels) is much larger than the binding momentum. 
In systems like $T_{bc}$ and $T_{bb}$, the nearby thresholds violate this scale separation. 
The inter-channel transitions act as long- and intermediate-range dynamic forces that cannot be legitimately integrated out. 
Thus, an explicit coupled-channel treatment is not just an improvement, but a theoretical necessity to analyse the amplitude structure.

In this work, with the calibrated OBE potential, we then perform a systematic search for possible molecular states in the \(D^{(*)}D^{(*)}\), \(\bar{B}^{(*)}\bar{B}^{(*)}\), and \(D^{(*)}\bar{B}^{(*)}\) systems with total spin \(J = 1\).
In this comprehensive framework, we find that the explicit inclusion of coupled-channel dynamics fundamentally transforms the nature of the interaction.
Through the comparison between two schemes, we explicitly demonstrate the transformation of single-channel bound states induced by coupled-channel dynamics.
For some states, these couplings provide additional attraction that deepens the binding. 
For others, the strong non-diagonal terms can fundamentally flip the interaction. 
The attraction predicted within the single-channel framework is replaced by a manifest effective repulsion.
This phenomenon is not merely a quantitative correction but a universal feature of coupled-channel systems, appearing in both the $T_{cc}$ and $T_{bb}$ sectors. 
This insight offers a more comprehensive perspective on the origin of hadronic molecules.

This paper is structured as follows. 
In Section~\ref{sec:formul}, we introduce the theoretical framework, including the effective potentials the complex scaling method and $T$-matrix pole analysis. Section~\ref{sec:coup} is devoted to the analysis of coupled-channel effects.
In Section~\ref{sec:num}, we conduct a detailed analysis of coupled-channel effects and perform parameter fitting under different schemes based on the experimental lineshape of the $T_{cc}$ state.
Using two illustrative examples, we systematically examine both single-channel and coupled-channel dynamics in the $DD^{*}$-$D^{*}D^{*}$ and $\bar{B}\bar{B}^{*}$-$\bar{B}^{*}\bar{B}^{*}$ systems. 
Finally, a brief summary is presented in Section~\ref{sec:sum}. 
Some calculation details are presented in the appendices.

\section{Formulation} \label{sec:formul}
\subsection{The effective potentials}\label{sec:eff}

Under the heavy quark limit and chiral symmetry, the effective Lagrangians are~\cite{Falk:1992cx,Cheng:1992xi,Casalbuoni:1996pg,Ding:2008gr,Sun:2011uh}:
\begin{eqnarray}
\mathcal{L}_{P^{(*)}P^{(*)}M} 
     &=&-i\frac{2g}{f_\pi}\varepsilon_{\alpha\mu\nu\sigma}
     v^\alpha P^{*\mu}_{b}\partial^\nu M_{ba}{P}^{*\sigma\dag}_{a}\ 
     +\ i \frac{2g}{f_\pi}\varepsilon_{\alpha\mu\nu\sigma} 
     v^\alpha\widetilde{P}^{*\mu\dag}_{a}\partial^\nu{}M_{ab}
     \widetilde{P}^{*\sigma}_{b} \qquad \quad  
    \nonumber\\
    &~&-\frac{2g}{f_\pi}(P_bP^{*\dag}_{a\sigma}+P^{*}_{b\sigma}
    P^{\dag}_{a})\partial^\sigma{} M_{ba}
    \ +\ \frac{2g}{f_\pi}(\widetilde{P}^{*\dag}_{a\sigma}\widetilde{P}_b+ \widetilde{P}^{\dag}_{a}\widetilde{P}^{*}_{b\sigma})\partial^\sigma M_{ab}, 
\label{lagrangian:p}
\end{eqnarray}
\begin{eqnarray}
\mathcal{L}_{P^{(*)}P^{(*)}V}
     &=& -\sqrt{2}\beta{}g_VP_b v\cdot\hat{\rho}_{ba}P_a^{\dag}
        -2\sqrt{2}\lambda{}g_V \varepsilon_{\sigma\mu\alpha\beta}
        v^\sigma(P^{}_bP^{*\mu\dag}_a +P_b^{*\mu}P^{\dag}_a)
        (\partial^\alpha{}\hat{\rho}^\beta)_{ba}
        \nonumber\\
    &~&+\sqrt{2}\beta g_V\widetilde{P}^{\dag}_a v\cdot\hat{\rho}_{ab} \widetilde{P}_b
       -2\sqrt{2}\lambda{}g_V\varepsilon_{\sigma\mu\alpha\beta}v^\sigma
       (\widetilde{P}^{*\mu\dag}_a\widetilde{P}^{}_b+\widetilde
       {P}^{\dag}_a\widetilde{P}_b^{*\mu})(\partial^\alpha{}\hat{\rho}^\beta)_{ab} 
        \nonumber\\
    &~&+\sqrt{2}\beta{}g_V P_b^{*}\cdot P^{*\dag}_a v\cdot\hat{\rho}_{ba}
       -i2\sqrt{2}\lambda{}g_V P^{*\mu}_b (\partial_\mu{}\hat{\rho}_\nu 
       -\partial_\nu{}\hat{\rho}_\mu)_{ba}P^{*\nu\dag}_a \nonumber\\
    &~&-\sqrt{2}\beta g_V\widetilde{P}^{*\dag}_a\cdot\widetilde{P}_b^{*} v\cdot\hat{\rho}_{ab}
        -i2\sqrt{2}\lambda{}g_V\widetilde{P}^{*\mu\dag}_a(\partial_\mu{} \hat{\rho}_\nu 
        -\partial_\nu{}\hat{\rho}_\mu)_{ab}\widetilde{P}^{*\nu}_b, 
\label{lagrangian:v}
\end{eqnarray}
where $P_a=(D^+,D^0,D_s^0) / (B^-,\bar{B}^0,\bar{B}_s^0)$ and $P^*_a=(D^{*-},D^{*0},D_s^{*0}) / (B^{*-},\bar{B}^{*0},\bar B_s^{*0})$ 
are the heavy meson fields. 
The coupling constant $g=0.57$ is determined from the strong decay $D^*\rightarrow D\pi$  ~\cite{ParticleDataGroup:2022pth,CLEO:2001foe,CLEO:2001sxb}. 
The constant \(g_V = 5.8\) is introduced to compare our parameters with other phenomenological studies~\cite{Liu:2008fh,Li:2012ss}. 
The parameters $\lambda$ and $\beta$ are the additional coupling constants.
The exchanged pseudoscalar meson and vector meson matrices, $M$ and
$\hat{\rho}^{\mu}$, are distinctly expressed as
\begin{eqnarray}
M=\left( \begin{array}{ccc}
   {\pi^0\over\sqrt{2}}+{\eta\over\sqrt{6}} & \pi^+  & K^+ \\
   \pi^- & -{\pi^0\over\sqrt{2}}+{\eta\over\sqrt{6}} & K^0 \\
    K^-  & \bar{K}^0                                 & -{2\over\sqrt{6}}\eta \\
\end{array}\right ), \quad
\hat{\rho}^{\mu}=\left(\begin{array}{ccc}
  {\rho^0\over\sqrt{2}}+{\omega\over\sqrt{2}}      & \rho^+ & K^{*+} \\
  \rho^-  & -{\rho^0\over \sqrt{2}}+{\omega \over \sqrt{2}} & K^{*0} \\
  K^{*-}  & \bar{K}^{*0}                                    & \phi   \\
\end{array}\right)^{\mu}.
\end{eqnarray}
We consider couple-channel calculations involving the exchange of $\pi, \rho, \omega$ mesons. 
The contributions of $\sigma$ and $\eta$ mesons are ignored because their contributions are tiny~\cite{Li:2012cs,Li:2012ss,Wang:2018jlv}.
The effective potential, which serves as the interaction kernels entering the coupled-channel Schrödinger Equation(SE) and Lippmann–Schwinger Equation(LSE), can be written as
\begin{equation}
	{\cal V} = V_{\pi}^{u}+V_{\pi}^{t}+V^{u}_{\rho/\omega}+V^{t}_{\rho/\omega}.
\label{eq:v}
\end{equation}
The explicit forms of the potential are given in Appendix~\ref{app:PP}, and those of the polarization vectors in Appendix~\ref{app:Polar}.
A form factor is introduced to ensure regularization and convergence of the potential, i.e.,
\begin{eqnarray}
   {\cal F}\left(p_{i},p_{f}\right) = \left(\frac{\Lambda^{2}}
	{\Lambda^{2}+p_{f}^{2}}\right)^{2}\left(\frac{\Lambda^{2}}
	{\Lambda^{2}+p_{i}^{2}}\right)^{2}, \quad \mathcal V\rightarrow \mathcal V {\cal F}\left(p_{i},p_{f}\right)\label{eq:ff}
    \end{eqnarray}
where $\Lambda$ is the cutoff parameter. 
The physical observables are insensitive to the specific choice of the cutoff parameter $\Lambda$, as its dependence can be reabsorbed into the coupling constant $\lambda$ and $\beta$.

\begin{table}
	\centering
	  \renewcommand\arraystretch{1.5} 
        \small
	\vspace{1pt}
	\label{tab:PPOBE}
		\begin{tabular}{c|ccc}
         \hline\hline
$I(J^P)$ & System & $\pi$  & $\rho / \omega$ \\
\hline \hline
\multirow{4}{*}{$0(1^+)$} & $PP^*-PP^*$ & $\frac{3}{2}V_{\pi}^{u}$ &  $\frac{3}{2}V_{\rho}^{u}-\frac{1}{2}V_{\omega}^{u}-\frac{3}{2}V_{\rho}^{t}+\frac{1}{2}V_{\omega}^{t}$ \\
\cline{2 - 4} 
 & $PP^*-P^*P^*$ & $\frac{3}{2\sqrt{2}}(V_{\pi}^{u}-V_{\pi}^{t})$ & $\frac{1}{\sqrt{2}}\left(\frac{3}{2}V_{\rho}^{u}(\lambda)-\frac{1}{2}V_{\omega}^{u}(\lambda)-\frac{3}{2}V_{\rho}^{t}(\lambda)+\frac{1}{2}V_{\omega}^{t}(\lambda)\right)$ \\
\cline{2 - 4} 
 & $P^*P^*-PP^*$ & $\frac{3}{2\sqrt{2}}(V_{\pi}^{u}-V_{\pi}^{t})$ & $\frac{1}{\sqrt{2}}\left(\frac{3}{2}V_{\rho}^{u}(\lambda)-\frac{1}{2}V_{\omega}^{u}(\lambda)-\frac{3}{2}V_{\rho}^{t}(\lambda)+\frac{1}{2}V_{\omega}^{t}(\lambda)\right)$ \\
\cline{2 - 4} 
 & $P^*P^*-P^*P^*$ & $V_{\pi}^{u}-\frac{1}{2}V_{\pi}^{t}$ & $V_{\rho}^{u}(\lambda)+V_{\rho}^{u}(\beta)-\frac{1}{2}(V_{\rho}^{t}(\lambda)+V_{\omega}^{t}(\lambda)-V_{\rho}^{t}(\beta)+V_{\omega}^{t}(\beta))$ \\
\hline
\multirow{4}{*}{$1(1^+)$} & $PP^*-PP^*$ & $\frac{1}{2}V_{\pi}^{u}$ & $\frac{1}{2}V_{\rho}^{u}+\frac{1}{2}V_{\omega}^{u}+\frac{1}{2}V_{\rho}^{t}+\frac{1}{2}V_{\omega}^{t}$ \\
\cline{2 - 4} 
 & $PP^*-P^*P^*$ & $\frac{1}{2\sqrt{2}}(V_{\pi}^{u}+V_{\pi}^{t})$ & $\frac{1}{\sqrt{2}}\left(\frac{1}{2}V_{\rho}^{u}(\lambda)+\frac{1}{2}V_{\omega}^{u}(\lambda)+\frac{1}{2}V_{\rho}^{t}(\lambda)+\frac{1}{2}V_{\omega}^{t}(\lambda)\right)$ \\
\cline{2 - 4} 
 & $P^*P^*-PP^*$ & $\frac{1}{2\sqrt{2}}(V_{\pi}^{u}+V_{\pi}^{t})$ & $\frac{1}{\sqrt{2}}\left(\frac{1}{2}V_{\rho}^{u}(\lambda)+\frac{1}{2}V_{\omega}^{u}(\lambda)+\frac{1}{2}V_{\rho}^{t}(\lambda)+\frac{1}{2}V_{\omega}^{t}(\lambda)\right)$ \\
\cline{2 - 4} 
 & $P^*P^*-P^*P^*$ & $V_{\pi}^{u}-\frac{1}{2}V_{\pi}^{t}$ & $V_{\rho}^{u}(\lambda)+V_{\rho}^{u}(\beta)-\frac{1}{2}(V_{\rho}^{t}(\lambda)+V_{\omega}^{t}(\lambda)-V_{\rho}^{t}(\beta)+V_{\omega}^{t}(\beta))$ \\
\hline \hline
\end{tabular}
	\caption{The $P^{(*)}P^{(*)}$ ($P=D, B$) interactions derived within the OBE model under the isospin symmetry limit. }
\label{tab:obe}
\end{table}

In the isospin symmetry limit, the effective potentials can be classified into $I=0$ and $I=1$ channels as summarized in Table \ref{tab:obe}, which explicitly demonstrates the possible coupled effects between different $P^{(*)}P^{(*)}$ channels with the same $I(J^{P})$ quantum numbers. 
However, in the present work, perform calculations in the physical basis, where the physical masses of
$D^{+}D^{*0}$ and $D^{0}D^{*+}$ are explicitly retained. 
As a result, the isospin symmetry is broken, and the different isospin components are dynamically mixed.

\subsection{Complex scaling method}
\label{sec:csm}
With the effective interactions constructed above, we will explore the pole structures of the coupled-channel dynamics, with the complex scaling method (CSM), which offers a unified framework for the bound states, scattering states and resonant states simultaneously.
In momentum space, the SE is formulated as 
\begin{eqnarray}
{\cal H}_0(p)\Phi_i(p)+\int_0^{+\infty} \frac{p^{\prime 2} d p^{\prime}}{(2 \pi)^3} {\cal V}_{ij}\left(p, p^{\prime}\right) \Phi_j\left(p^{\prime}\right)=E\Phi_i(p),
\end{eqnarray}
where ${\cal H}_0=\sqrt{m_{i,1}^{2}+p^{2}}+\sqrt{m_{i,2}^{2}+p^{2}}$ is the kinematic energy with $m_{i,1/2}$ the masses of the constituents in the $i$-th channel and ${\cal V}$ is the effective potential defined in Eq.~\eqref{eq:ff}. 
By introducing a complex rotation  operator \( U(\theta) \), the coordinate \( r \) and its  momentum \( p \) are transformed as:  
\begin{eqnarray}
U(\theta)r = re^{i\theta}, \quad U(\theta)p = pe^{-i\theta}. 
\label{eq:CSM}
\end{eqnarray}
Here $\theta$ is a real positive scaling angle. 
Under this transformation, the Hamiltonian and the SE are correspondingly transformed, leading to the complex-scaled SE
\begin{eqnarray}
{\cal H}_0({p}e^{-i\theta}) \Phi_i({p}e^{-i\theta})+e^{-3i\theta}\int_{0}^{+\infty} \frac{{p}^{\prime 2} d {p}^{\prime}}{(2 \pi)^3} {\cal V}_{ij}\left({p}e^{-i\theta}, {p}^{\prime}e^{-i\theta}\right) \Phi_j\left({p}^{\prime}e^{-i\theta}\right)=E_{\theta}\Phi_i({p}e^{-i\theta}),\label{csm-eq}
\end{eqnarray}
Within this framework,
the bound states are located on the negative real axis in the energy plane and do not change under the rotation angle \(\theta\).
Resonant states, whose wave functions are not square-integrable in conventional approaches, become normalizable once the rotation angle \(\theta\) exceeds a critical value (e.g., $\theta>\frac{1}{2} {tan}^{-1}(\Gamma/2 E_r)$ for a resonant pole $E_{\text{pole}}=E_r-i\frac{\Gamma}{2}$). 
This normalization enables treating resonant states analogously to bound states. 
Meanwhile, the continuum spectrum is rotated along lines, with $\arg(E)=-2\theta$, providing a clear separation between resonant poles and scattering states.
In practical calculations, $\theta$ should not be excessively large to alter the damping behavior of the potential at large momenta. 
Further details on CSM can be found in Refs.~\cite{Myo:2014ypa,
Myo:2020rni,Aoyama:2006,moiseyev1998quantum,Chen:2023eri}.

The CSM provides a direct and efficient way to access the bound and resonant poles spectrum without requiring a numerical search in the complex energy plane. 
However, virtual states, which are typically located on the second Riemann sheet below threshold, are difficult to resolve within the CSM framework. 
On the other hand, the  $T$-matrix approach offers a complementary description in terms of scattering amplitudes, which can be directly connected to experimental observables such as line shapes. 
It also allows for a reliable identification of virtual states through analytic continuation. 
In this work, we employ both methods in a complementary manner.


\subsection{$T$-matrix pole analysis}\label{sec:tmax}
As a complementary approach to the CSM, the pole structures can be  investigated through the scattering amplitude. 
The scattering amplitude $T$, is related to the effective potential $\mathcal{V}$ via the LSE, as follows: 
\begin{eqnarray}
T = \mathcal{V} + \mathcal{V} G T.
\end{eqnarray}
where $G$ is the diagonal matrix of two-body propagators. 
The LSE is a non-Hermitian integral equation that contains the full dynamics of particle interactions. 
In matrix form, the general coupled-channel LSE becomes~\cite{Nakamura:2015rta, Wu:2012md, Wu:2014vma, Hao:2025ukk}:
\begin{eqnarray}
T_{ij}\left(k, k^{\prime} ; E\right)= & \mathcal{V}_{i,j}\left(k, k^{\prime} ; E\right)+\sum_{i^{\prime}} \int_{0}^{\infty} q^{2} d q \mathcal{V}_{i, i^{\prime}}(k, q ; E) G_{i^{\prime}}(q ; E) T_{i^{\prime}, j}\left(q, k^{\prime} ; E\right).
\end{eqnarray}
Here, $k$, $k^{\prime}$ and $q$ are the relative momenta of the initial, final and intermediate states respectively, and \( i,j \) label different channels.
The scattering energy is denoted by $E$. 
For the $i^{\prime}$-th channel, its propagator $G_{i^{\prime}}(q ; E)$ is defined as 
\begin{eqnarray}
G_{i^{\prime}}(q ; E)=\frac{1}{E-E_{i^{\prime},a}(q)-E_{i^{\prime},b}(q)+i \varepsilon}, 
\end{eqnarray}
where $E_{i^{\prime},a/b}(q)=\sqrt{m_{i^{\prime},a/b}^{2}+q^{2}}$ represents the kinematic energies of the intermediate mesons $a$ or $b$. 

The positions of poles in the complex energy plane correspond to different types of states (bound, virtual, and resonant) depending on the strength of the interaction.
We can identify the states by solving the equation $\text{det}[1 - V G(E_{\text{pole}})] = 0$.
When the interaction is strong and attractive enough to generate a bound state, the pole would be located below the two-hadron threshold on the first (physical) Riemann sheets (RS).
If the attraction is insufficient, the pole as a virtual state can be found in the second RS, still below the threshold. 
And when the interaction is even weaker, the poles associated with resonant states will appear above the threshold on the second RS.
In coupled-channel scattering, multiple channels are involved and the LSE is extended to account for interactions between these channels, leading to rich pole dynamics. 
The $T$-matrix formalism thus provides a direct link between the pole structure and experimentally observable quantities such as line shapes, making it a powerful tool for investigating the nature of exotic states.

\section{Analysis of coupled-channel effects}\label{sec:coup}
The coupled-channel dynamics play a crucial role in understanding near-threshold heavy-hadron systems. 
Within the HQSS, the effective interaction strengths in the $PP^*$ and $P^*P^*$ channels are of the same order, and the mass splitting $\Delta M = M_{P^{*}} - M_{P} \sim \Lambda_{\text{QCD}}^{2}/m_{Q}$ is suppressed in the heavy-quark limit. 
As a result, the $PP^{*}$ and $P^{*}P^{*}$ systems exhibit similar dynamics and are naturally coupled.

In the single-channel scheme, the short-range couplings generally absorb not only genuine contact interactions but also the contributions from omitted coupled channels. 
Consequently, such parameters can successfully reproduce the observed line shape in a specific channel.
However, this also implies that the extracted couplings are not purely intrinsic low-energy constants but depend on how the underlying dynamics are organized. 
In particular, their directly use as an universal low-energy constants for other HQSS-related systems may become unreliable.

When the coupled channels are treated explicitly, part of the dynamics previously hidden in the short-range interactions is resolved dynamically. 
For the $PP^*-P^*P^*$ system,  one constructs a coupled-channel potential matrix: \begin{eqnarray}
\mathcal{V}=\begin{pmatrix}v_{PP^* \to PP^*}&v_{PP^* \to P^*P^*}\\v_{P^*P^* \to PP^*}&v_{P^*P^* \to P^*P^*}\end{pmatrix}.
\end{eqnarray}
Here, the off-diagonal coupling effects introduce additional ultraviolet contributions, leading to a stronger cutoff dependence. 
In a consistent framework, such dependence should be absorbed by a redefinition of the coupling constants. 
However, if the couplings are fixed beforehand in a single-channel analysis, they cannot fully absorb the newly generated cutoff dependence. 
Thus, this can significantly shift the pole position. 
This mechanism explains the large discrepancies observed between single-channel and coupled-channel calculations. 
For example, using the parameters fitted in the single-channel $DD^*$ analysis that reproduce the $T_{cc}$ binding energy of $393$ keV, we find that the inclusion of $D^*D^*$ channel shifts the bound state to $53$ keV. 
This difference does not arise from an inconsistency, but reflects different organizations of the same low-energy dynamics.

In addition, explicit coupled-channel dynamics naturally generate threshold effects such as cusp structures.
Lattice QCD studies confirm a robust $S$-wave coupling between the $DD^*$ and $D^*D^*$ channels, which generates a clear cusp in the $DD^*$ amplitude at the $D^*D^*$ threshold~\cite{Whyte:2024ihh}.
While the single-channel scheme provides an effective description of a specific process, an explicit coupled-channel treatment is also needed for a dynamical interpretation of the state and for reliable HQSS-based extrapolations to other systems.
In the following, taking the $T_{cc}$ system as an explicit example, we adopt two distinct strategies to fit the experimentally observed lineshape in order to systematically investigate the impact of the coupled-channel effects. 
%
%
%

\section{Results and discussions}\label{sec:num}

\subsection{Fitting schemes}\label{sec:fit}

\begin{figure}[htp]
	\centering
	\begin{tabular}{cc}
		\includegraphics[width=0.42\textwidth]{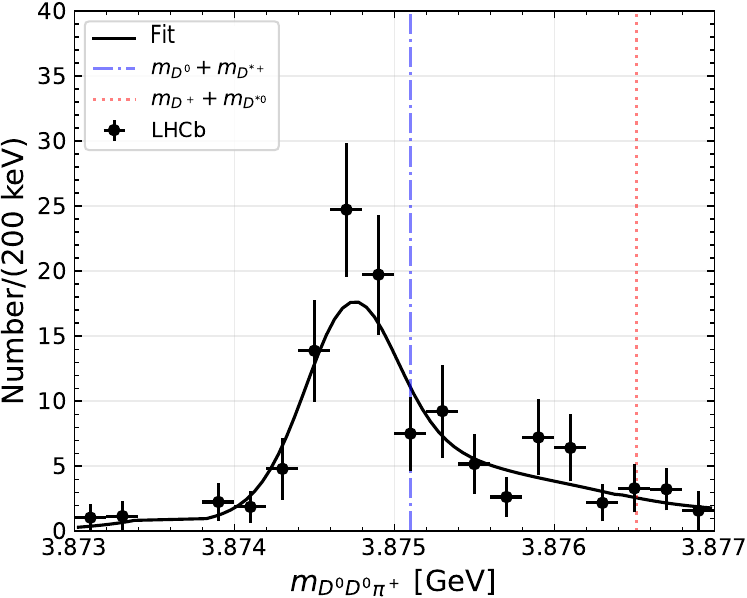} &  
		\includegraphics[width=0.42\textwidth]{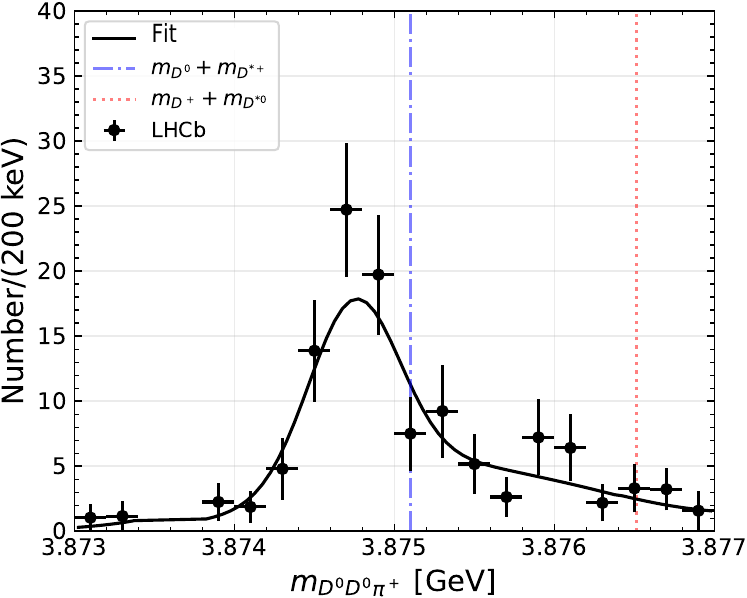} 
    \end{tabular}
    \begin{tabular}{c}
		\includegraphics[width=0.44\textwidth]{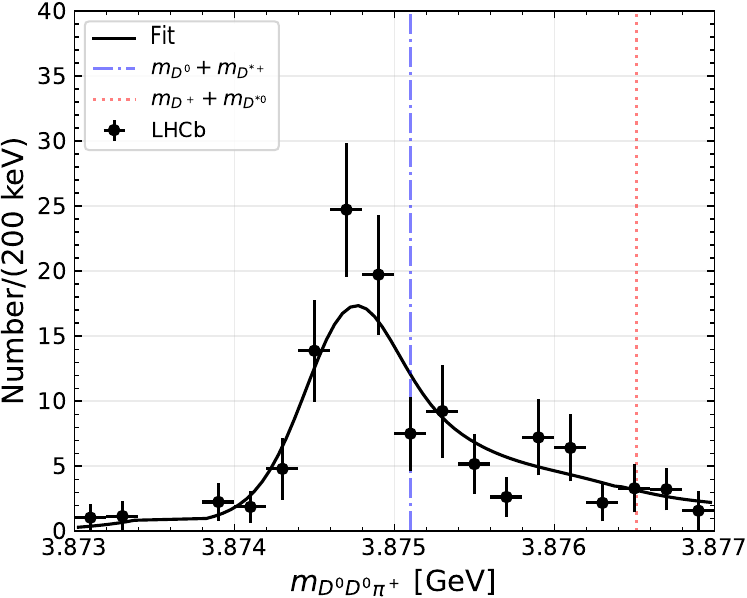} 
    \end{tabular}
    \caption{ 
    Fitted lineshapes of the $T_{cc}^{+}$ in the $D^0D^0\pi^+$  invariant mass spectrum based on the coupled-channel framework (Scheme II). The lineshapes are convolved with the experimental energy resolution for three different cutoff values: $\Lambda = 0.8$, $1.0$, and $1.2$ GeV. 
    }
    \label{fig:fit}
\end{figure}

The cutoff reflects the composite structure of hadrons and off-shell effects, and adjusting the cutoff primarily affects the $T$-matrix by compensating for UV divergence. 
The effective potential of vector meson exchange in Eq.~\eqref{eq:ff} contains two unknown coupling constants: $\lambda$ and $\beta$. 
To extract the interactions between \( D \) and \( D^* \), we fit the experimental lineshape of the \( T_{cc}^+ \) state for three cutoff values $\Lambda = 0.8$ GeV, $1.0$ GeV and $1.2$ GeV with two different schemes:
\begin{itemize}
	\item Scheme I (single-channel): only the $DD^*$ channel is considered. 
	\item Scheme II (coupled-channel systems): The $DD^*$–$D^*D^*$ coupled-channel interactions are included explicitly. 
    The diagonal interactions are the same as in Scheme I, while the off-diagonal potentials describe the coupling between the two channels, induced by the same set of light-meson exchanges.
\end{itemize}
Fit results of the Scheme II are shown in Fig.~\ref{fig:fit}, with the best-fit parameters listed in Table~\ref{tab:fit}, while the corresponding results for Scheme I can be found in our work~\cite{Wang:2023ovj}. 
Both of the schemes can well describe the experimental data.
With heavy quark flavor symmetry, these parameters can be further applied to $\bar{B}^{(*)}\bar{B}^{(*)}$ and $D^{(*)}\bar{B}^{(*)}$ system. 

\begin{table}
	\centering
	  \renewcommand\arraystretch{1.5} 
        \setlength{\tabcolsep}{2mm}   
		\begin{tabular}{ccccc}
         \hline\hline
			$\Lambda$ (GeV) & Scheme I & $\chi^{2}/\mathrm{d.o.f.}$ & Scheme II & $\chi^{2}/\mathrm{d.o.f.}$                         \\ \hline\hline
			0.8    & $\lambda=0.890/\mathrm{GeV}, \beta=0.810$ &  0.78& $\lambda=0.490/\mathrm{GeV}, \beta=0.777$ &   0.68 \\ 
			1.0    & $\lambda=0.683/\mathrm{GeV}, \beta=0.687$ &   0.76 & $\lambda=0.220/\mathrm{GeV}, \beta=0.630$ &   0.66 \\ 
			1.2    & $\lambda=0.587/\mathrm{GeV}, \beta=0.550$ &   0.78 & $\lambda=0.104/\mathrm{GeV}, \beta=0.332$ &   0.80  \\ 
        \hline\hline
		\end{tabular}
	\caption{The best-fitted values of parameters $\lambda$ and  $\beta$ values together with the corresponding \(\chi^2\)/d.o.f. for fitting Scheme I (single-channel) and Scheme II (coupled-channel) under different cutoff values \(\Lambda = 0.8, 1.0, 1.2\,\mathrm{GeV}\).}
    \label{tab:fit}
\end{table}

\subsection{Single-channel systems}\label{sec:1chan}

\subsubsection{$DD^{*}$ and $D^*D^{*}$ }
\begin{figure}[htp]
	\centering
	\begin{tabular}{cc}
		\includegraphics[width=0.45\textwidth]{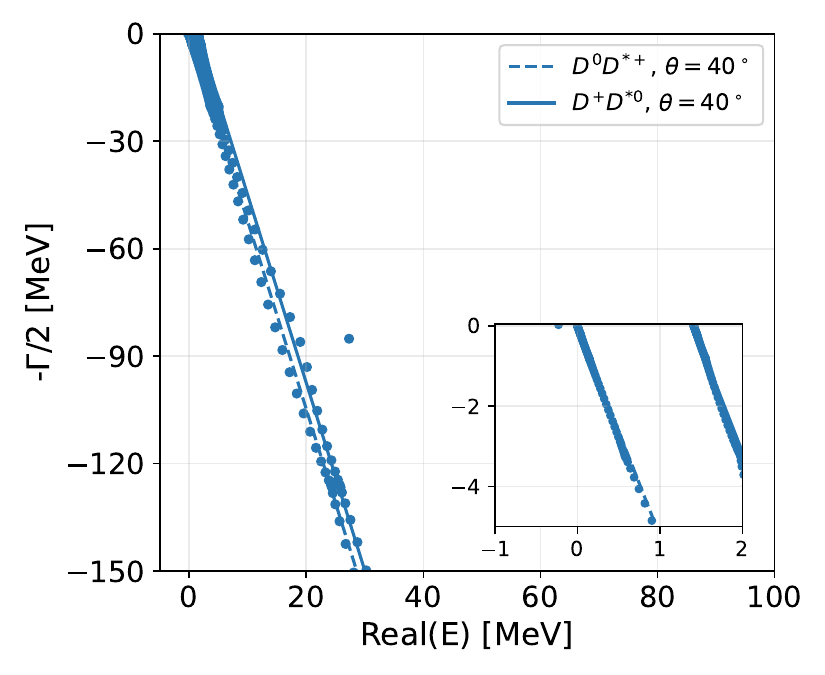} &  
		\includegraphics[width=0.45\textwidth]{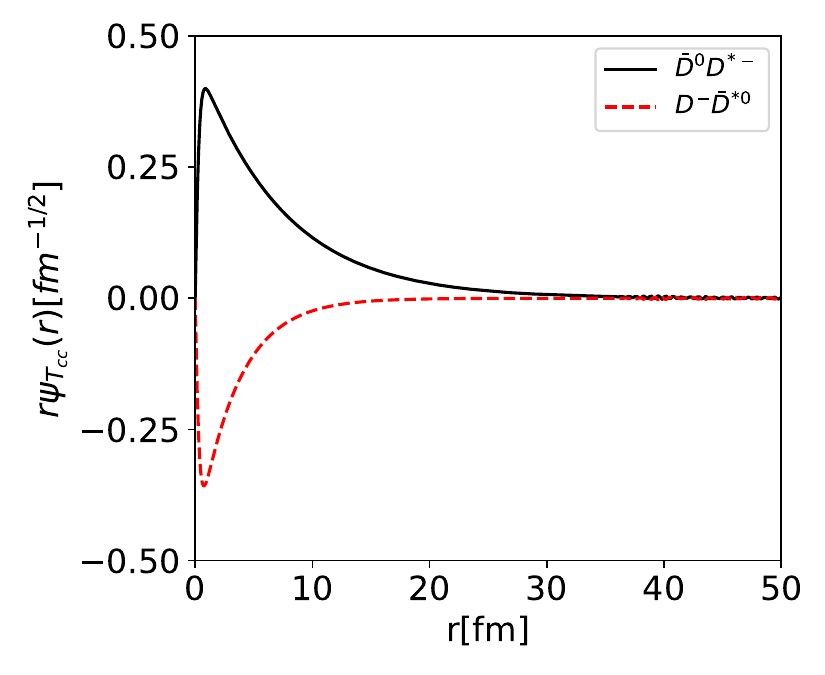} \\  
		\multicolumn{2}{c}{\small{(a) $DD^{*}$} system} \\  
		\includegraphics[width=0.45\textwidth]{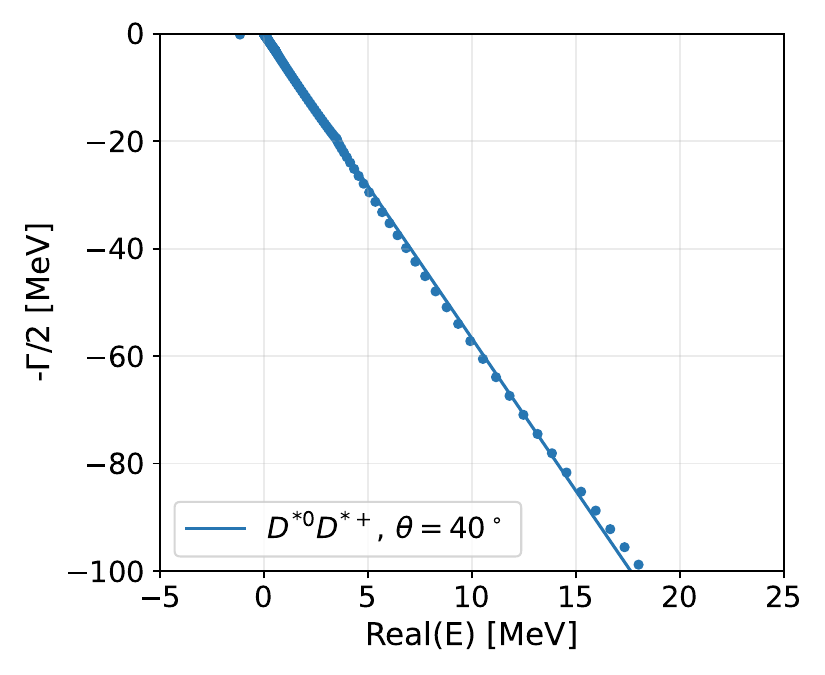} &  
		\includegraphics[width=0.45\textwidth]{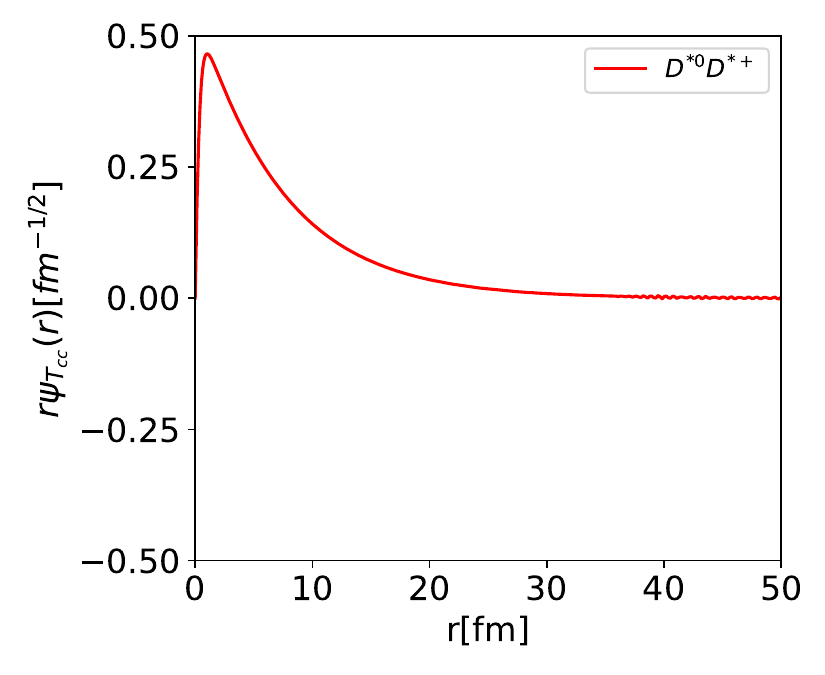} \\ 
		\multicolumn{2}{c}{\small{(b) $D^{*}D^{*}$} system} 
	\end{tabular}
	\caption{The complex eigenvalues (left) and the corresponding coordinate-space wave functions (right) of the bound states  for the $DD^{*}$ (a) and $D^{*}D^{*}$ (b) systems at $\Lambda = 1.0\,\mathrm{GeV}$.}
        \label{fig:D(s)Ds}
\end{figure}

We summarize the results for $DD^{*}$ and $D^*D^{*}$ in Table~\ref{tab:Pole}. 
For the $DD^*$ channel, the main results have been reported in  Ref.~\cite{Wang:2023ovj}, and are briefly recalled here for completeness.
The $T_{cc}^+$ can be well described as a $D^*D$ molecular state with quantum numbers $J^P = 1^+$,  exhibiting a shallow binding energy of $-393.0$ keV and a width of about $70$ keV.
The isospin analysis shows that the state is dominated by the isoscalar component ($95.8\%$), with only a small isovector admixture($4.2\%$). 
Its large root mean square radius $\sqrt{\langle r^2\rangle}\approx 4.7$ fm, together with the long-tail of the wave function are shown in Fig.~\ref{fig:D(s)Ds} (a). 
Although the mass splitting between the two charged channels is rather small of $1.4$ MeV, the even extremely weak binding energy around $400$ kev leads to the visible isospin symmetry breaking.
The dependence of these states with different cutoff values is presented in Appendix~\ref{app:Cutoff}.

Notably, LHCb measurements show no significant structures in the $D^{+} D^{0} \pi^{+}$ and $D^{+} D^{+} \pi^{0}$ channels up to 4 GeV. 
Since these channels are dominated by the $I=1$ component, the absence supports suggest that the observed $T_{cc}^{+}$ is an isoscalar ($I=0$) state. 
Further support comes from recent lattice QCD studies, which reveal a repulsive $I=1$ $D D^{*}$ interaction that prevents the formation of an isovector state, in contrast to the attractive forces found in the $I=0$ channel~\cite{Meng:2024kkp,Chen:2022vpo}.

Despite no signals from experiments and lattice QCD, the existence of isospin-triplet doubly charmed tetraquarks is predicted in various theoretical studies~\cite{Ader:1981db,Luo:2017eub,Albaladejo:2021vln,Giron:2021sla}.
The absence of experimental signals may be attributed to suppressed production rates in proton-proton collisions, or line-shape distortions from coupled-channel dynamics near threshold. 
Consequently, high-statistics Dalitz plot analyses~\cite{Shi:2022slq} are proposed to search for these elusive isovector states. 
Heavy-ion collisions have been proposed as a more favorable environment for generating doubly charmed tetraquarks, especially the elusive isospin-triplet states, owing to the high charm-quark density and distinct production mechanisms relative to those in $pp$ collisions~\cite{Hu:2021gdg}.

Motivated by these possibilities, we employ the CSM to explore the $I=1$ sector in the higher-energy region.
Our analysis reveals a distinct resonance pole located at $E_r \approx 23.0$ MeV above the $D^{*+}D^0$ threshold, with a significant width of $\Gamma = 170.3$ MeV. 
Wave function analysis confirms that this state is dominated by the $I=1$ component. 
This structure was not reported in previous works~\cite{Wang:2023ovj}, because the study did not extend into the higher-energy range. 
Such a broad resonance in the isovector channel escapes detection by blending into the background as a diffuse signal. 
This discovery provides a self-consistent explanation for the null experimental results.

A resonance with mass $3902.4$ MeV and width $170$ MeV also exists in the $I=1$ $DD^*$ system. 
The resonance masses and widths for different cutoff values are discussed in Appendix~\ref{app:Cutoff}.
The same result is confirmed by the $T$-matrix calculation on the second Riemann sheet.
Due to its large width, this resonance is highly unstable and its signal is spread over a broad energy region, making it difficult to distinguish from the smooth background in experiments.

We further studied the \(D^* D^*\) systems. 
For the \( S \)-wave \( D^*D^* \) configurations, the allowed quantum numbers are \( I(J^P) = 0(1^+) \), \( 1(0^+) \) or \( 1(2^+) \). 
For the $J^P=1^+ $ channel, we find only one bound state with isospin $I=0$, as summarized in Table~\ref{tab:Pole}. 
At $\Lambda=1.0$ GeV, the binding energy is $1.976 $ MeV and the root-mean-square radius is $2.47$ fm.
Fig.~\ref{fig:D(s)Ds} (b) shows the pole positions and the corresponding coordinate-space wave functions. 
The long-range tail of the wave function also indicates a molecular candidate dominated by the \( D^*D^* \) interaction. 
Such behavior is consistent with HQSS, which implies similar interaction patterns in the \(D^*D\) and \(D^*D^*\) systems in the heavy-quark limit.

\xiu{\begin{table}[htbp]
  \centering
  \renewcommand\arraystretch{1.5}
  \setlength{\tabcolsep}{5.0mm}   
  \vspace{1pt}
  \begin{tabular}{l |c c c c c} \hline \hline	
    System & State & Pole position (MeV) & $\sqrt{\langle r^2\rangle}$ & \\  \hline \hline			
    $DD^*$ & Bound state & 3876.1($\Delta E_B$ = 393 KeV) & 4.7 fm & \\ 
           & Resonance & 3902.4$-\dfrac{i}{2}$170.3 & & \\ \hline	
    $D^*D^*$ & Bound state & 4015.1($\Delta E_B$ = 1.976 MeV) & 2.47 fm & \\ \hline	
    $\bar{B}\bar{B}^*$ & Bound state & 10560.1($\Delta E_B$ = 43.9 MeV) & 0.61 fm & \\ 	
           & Resonance &  10621.2$-\dfrac{i}{2}$26.4 & & \\ \hline	
    $\bar{B}^*\bar{B}^*$ & Bound state & 10604.3($\Delta E_B$ = 45.1 MeV) & 0.60 fm & \\ \hline	
    $D\bar{B}^*$ & Virtual state & 7144.3($\Delta E_V$ = 45.2 MeV) & &  \\ 
          & Resonance & 7232.0$-\dfrac{i}{2}$298.1 &  &  \\ \hline	
    $\bar{B}D^*$ & Virtual state & 7247.2($\Delta E_V$ = 39.3 MeV) &  & \\ 
          & Resonance & 7330.3$-\dfrac{i}{2}$283.8 &  &  \\ \hline	
    $D^*\bar{B}^*$ & Virtual state & 7292.4($\Delta E_V$ = 39.1 MeV) &  & \\ 
          & Resonance & 7375.6$-\dfrac{i}{2}$283.1 &  & \\ \hline \hline		
  \end{tabular}
  \caption{Pole positions of the $D^{(*)}D^*$, $B^{(*)}\bar{B}^{(*)}$, and $D^{(*)}\bar{B}^{(*)}$ systems with the cutoff parameter $\Lambda = 1.0$ GeV. Here, $\Delta E_B$ denotes the binding energy for bound states, $\Delta E_V$ denotes the energy of the virtual state below the threshold, resonances are listed with their decay widths, and $\sqrt{\langle r^2\rangle}$ represents the root-mean-square radius of the corresponding bound state.}
  \label{tab:Pole}
\end{table}
}

\subsubsection{$\bar{B}\bar{B}^{*}$ and $\bar{B}^{*}\bar{B}^{*}$}

The $\bar{B}^{(*)}\bar{B}^{(*)}$ systems were previously studied in our work~\cite{Ren:2023pip} within a single-channel framework, where the $\bar{B}\bar{B}^*$ and $\bar{B}^*\bar{B}^*$ channels were treated independently. 
The corresponding results are summarized in Table~\ref{tab:Pole} for reference.
In the $\bar{B}\bar{B}^*$ system, a deeply bound state with $J^P = 1^+$ is found. 
At $\Lambda=1.0$ GeV, the binding energy is about 44 MeV below the $\bar{B}\bar{B}^*$ threshold, and the state is nearly a pure isoscalar with a dominant $I=0$ component of $99.9\%$.
In addition, a resonant state with almost $I=1$ component is identified at a mass of $10621$ MeV with a width of $26.4$ MeV.  
For the $\bar{B}^*\bar{B}^*$ system, only one bound state with $I(J^P)=0(1^+)$ is found, with a binding energy of 45.1 MeV.  
The corresponding $I=1$ resonance does not appear due to the absence of an allowed configuration. 
Compared to the charm sector, the $\bar{B}^{(*)}\bar{B}^{(*)}$ systems exhibit significantly stronger binding, resulting in more compact wave functions and reduced sensitivity to long-range dynamics. 
Owing to the smaller mass splitting in the bottom sector, the $\bar{B}\bar{B}^*$ and $\bar{B}^*\bar{B}^*$ systems exhibit improved heavy-quark spin symmetry (HQSS) and isospin symmetry.

\subsubsection{$D\bar{B}^{*}$, $D^{*}\bar{B}$ and $D^{*}\bar{B}^{*}$}

\begin{figure}[htp]
	\centering
	\begin{tabular}{cc}
		\includegraphics[width=0.42\textwidth]{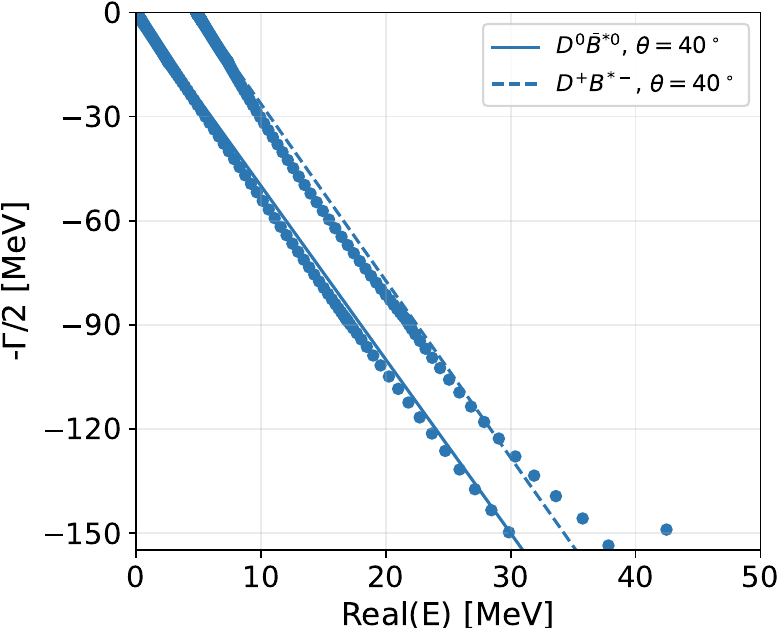} &  
		\includegraphics[width=0.42\textwidth]{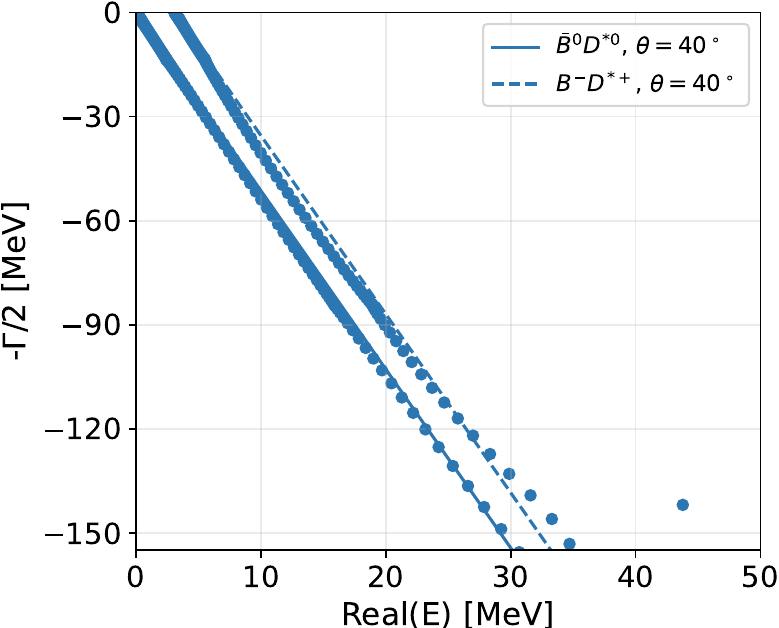} 
    \end{tabular}
		\includegraphics[width=0.42\textwidth]{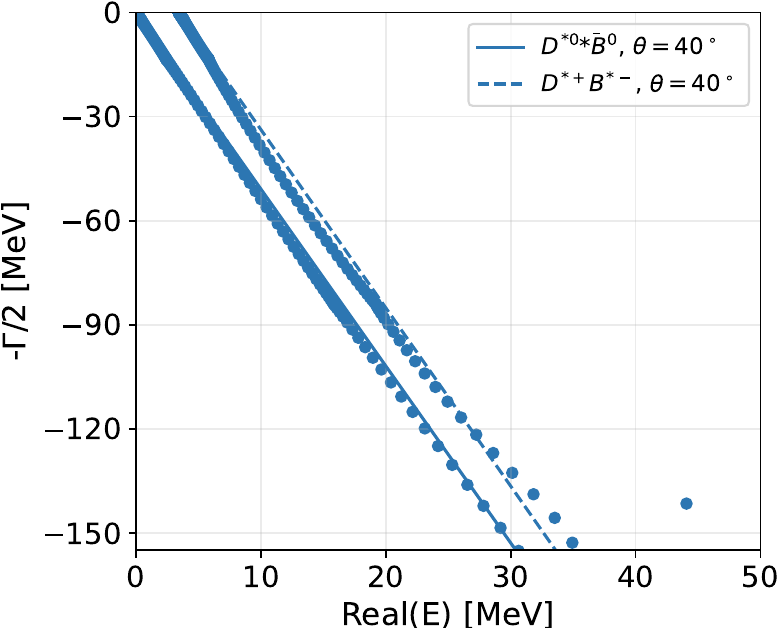} \\ 
	\caption{The complex eigenvalues obtained with CSM at \(\Lambda=1.0\,\mathrm{GeV}\) are presented in three panels for different channels: the top-left panel for \(D^{0}\bar{B}^{*0}\) and \(D^{+}\bar{B}^{*-}\), the top-right for the \(\bar{B}^{0}D^{*0}\) and \(\bar{B}^{-}D^{*+}\) channels, and the bottom-center for \(D^{*0}\bar{B}^{*0}\) and \(D^{*+}\bar{B}^{*-}\).}
    \label{fig:Tbc_res}
\end{figure}

We now discuss the $S$-wave single-channel interactions in the $D\bar{B}^{*}$, $D^{*}\bar{B}$ and $D^{*}\bar{B}^{*}$ systems.
The \(D^{(*)}\bar{B}^{(*)}\) systems, exhibit a mixed heavy quark flavor configuration. 
In the single-channel framework, there are only the \(t\)-channel contributions, while the $u$-channel contribution is absent. 
For the $D\bar{B}^{*}$ and $D^{*}\bar{B}$ systems, only vector-meson exchange contributes, whereas for the $D^{*}\bar{B}^{*}$ system both vector-meson and one-pion exchanges are present. 
Compared to the $D^{(*)}\bar{D}^{(*)}$ and $B^{(*)}\bar{B}^{(*)}$ systems, the absence of the long-range contribution results in a much weaker overall attraction. 
As a consequence, no bound states are obtained and only virtual states and resonances are obtained. 
The numerical results of pole positions are summarized in Table~\ref{tab:Pole}, and the corresponding distributions in the complex energy plane at $\Lambda=1.0$ GeV are shown in Fig.~\ref{fig:Tbc_res}.
%
%

For the $D\bar{B}^{*}$, a virtual state with $J^P = 1^+$ is located at $E_{v}=45.2$ MeV below the $D^{0}\bar{B}^{*0}$ threshold, appearing on the second RS of $D^{0}\bar{B}^{*0}$ channel and the first RS of $D^{+}\bar{B}^{*-}$ channel. 
%
Using the CSM, we also identify a $1(1^{+})$ resonance state above the threshold, located at $M = 7264.4$ MeV with a broad width of $\Gamma = 301.2$ MeV. 

For the $D^*\bar{B}$ and $D^*\bar{B}^*$ systems, we find that both exhibit a virtual state and a resonance with similar pole positions. 
This similar behavior is driven by the well-maintained heavy-quark symmetry between the $\bar{B}$ and $\bar{B}^*$ mesons.
Furthermore, we explore the cutoff $\Lambda$ dependence of these states. 
Taking the $D\bar{B}^*$ system as an example, when the cutoff $\Lambda$ decreases from $1.0$ GeV to $0.8$ GeV, the $\Delta E_v$ for the $D\bar{B}^*$ virtual state significantly shifts from $45.2$ MeV to $22.25$ MeV, moving closer to the threshold. 

We emphasize that these results are obtained within a single-channel framework using parameter set I. 
The corresponding coupling constants cannot be directly applied to coupled-channel systems, such as  $\bar{B}\bar{B}^*- \bar{B}^* \bar{B}^*$ system, which would otherwise lead to an overestimation of the binding energy \cite{Ren:2023pip}. 
Therefore, in the following,  we adopted the refitting procedure of Scheme II described in Subsec.~\ref{sec:fit}. 

\subsection{Coupled-channel systems}\label{sec:2chan}

\subsubsection{$DD^{*}-D^{*}D^{*}$ }
\begin{figure}[htp]
	\centering
	\begin{tabular}{cc}
			\includegraphics[width=0.48\textwidth]{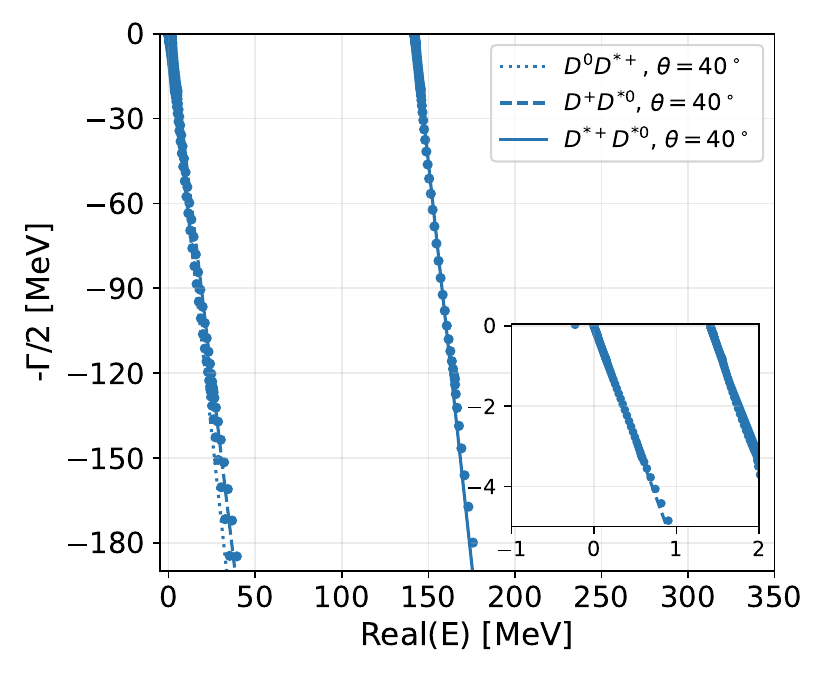}&
			\includegraphics[width=0.48\textwidth]{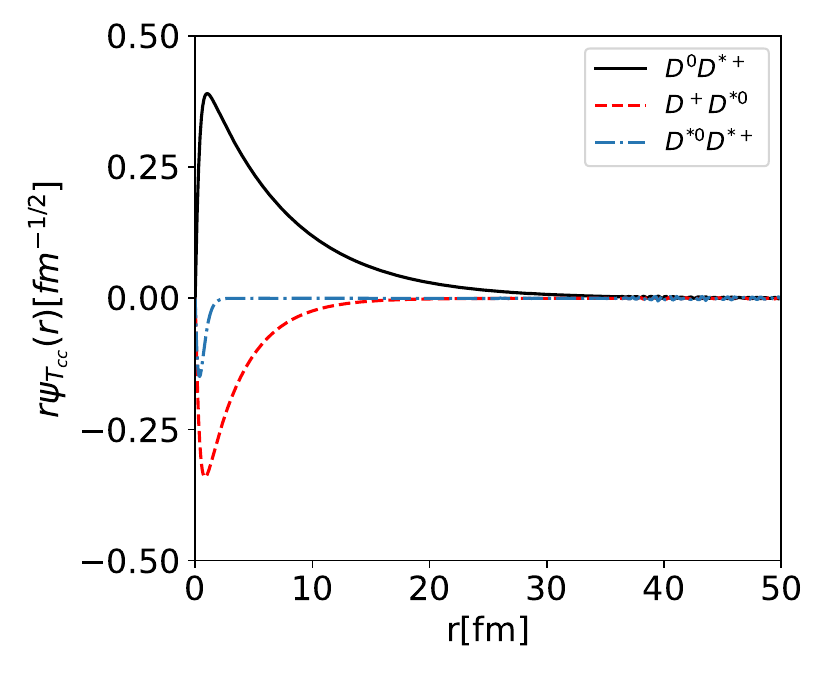}
		\end{tabular}
	\caption{The complex eigenvalues (left) and wave functions (right) of the three components with cutoff $\Lambda=1.0$ GeV for the $D D^*$-$D^*D^*$ system.}
    \label{fig:DDs_DsDs}
\end{figure}

We first investigate the $DD^{*}$-$D^{*}D^{*}$ system within the coupled-channel framework. 
After refitting the experimental $T_{cc}$ lineshape using Scheme II, we obtain the parameters $\lambda_{\Lambda=1.0} = 0.22 \, \text{GeV}^{-1}$ and $\beta_{\Lambda=1.0} = 0.63$. 
In the $J^P = 1^+$ sector, we find only a single bound state with a binding energy of $381.1$ keV, which can be identified as the $T_{cc}^+$ candidate. 
No additional bound or resonant states are observed. 
The pole structures and the corresponding wave functions at $\Lambda=1.0$ GeV are shown in figure~\ref{fig:DDs_DsDs}.
The properties of the bound state, including the root mean square radius and the proportions of the $D^{+} D^{*0}$, $D^0 D^{*+}$ and $D^{*+} D^{*0}$ channels are  summarized in Table~\ref{tab:com_coup_channels}.
The bound state is dominated by the $DD^{*}$ component with a probability $98.7\%$, consistent with Refs.~\cite{Li:2012ss,Asanuma:2023atv}.
This dominance can be attributed to the large mass gap ($\sim140$) MeV between the $DD^{*}$ and $D^{*}D^{*}$ thresholds. 
This result differs significantly from the single-channel scenario, an isoscalar $D^{*}D^{*}$ bound state and an isovector-dominated $DD^{*}$ state predicted in the Shechem I are both disappear once the coupled-channel included.

The underlying mechanism can be better understood as follows.
In the coupled-channel framework, part of the short-range dynamics in $T^+_{cc}$ is effectively encoded in the channel coupling, leading to a reduction of the effective attraction in each individual channel. 
This is reflected in the smaller value of the coupling constant $\lambda$ and $\rho$ obtained from the refit, which control the strength of vector-meson exchange in $DD^{*}$ and $D^{*}D^{*}$ channels. 
As a result, when the two channels are artificially decoupled but still using the parameters of Scheme II, both $DD^{*}$ and $D^{*}D^{*}$ systems only support virtual states. 
The $DD^{*}$ virtual state lies $1.76$ MeV below the $D^{0}D^{*+}$ threshold, while the $D^{*}D^{*}$ virtual state is also found $1.76$ MeV below $D^{*0}D^{*+}$ threshold.
These two virtual states correspond to the isoscalar-dominated $DD^*$ and isoscalar $D^{*}D^{*}$ bound states previously identified in single-channel scenario. 

Once the channel coupling is switched on, the off-diagonal interaction plays a dual role. 
For the $DD^{*}$ channel, the coupling to $D^{*}D^{*}$ provides an additional attractive contribution, which promotes the virtual state into a shallow bound state, identified as $T_{cc}^+$. 
In contrast, for the $D^{*}D^{*}$ channel, the coupling to $DD^{*}$ induces an effective repulsion, which pushes the virtual state away from the threshold and prevents the formation of a bound or resonant state within the energy region shown in Fig. \ref{fig:DDs_DsDs}. 
If the searched energy range is extended, a pole can be found at a much far away position, located about $72.37$ MeV above the lowest $DD^*$ threshold, with a quite large imaginary part $\frac{\Gamma}{2}\sim 400$ MeV. 
Such a pole lies far from the physical region and is therefore not expected to correspond to an experimentally observable state.
This explains why only a single bound state survives in the coupled system.

Another important consequence is the disappearance of the isovector-dominated $DD^*$ configuration.
This channel is expected to be largely decoupled from the $D^{*}D^{*}$ channel, since the latter only supports the $I=0$ configuration.
The absence can be traced back to the relative reduction of the effective couplings in Scheme II,  where the decrease of $\lambda$ (associated with the $u$-channel) is more significant than that of $\rho$ (associated with the $t$-channel).
As a result, in the $I=1$ $DD^{*}$ channel, the attractive $u$-channel contribution is overcome by the repulsive $t$-channel contribution, leading to an overall repulsive interaction, i.e., $V^{(I=1)}(DD^{*}) > 0$. 
As discussed in Ref.~\cite{Hao:2025ukk,NUSSENZVEIG1959499}, even in the presence of a repulsive interaction, a pole may still appear as a subthreshold broad pole located on an unphysical Riemann sheet, provided that the interaction is not too strong compared to the kinetic energy. 
Such a pole is typically located below the threshold but exhibits a large imaginary part, contributing as a smooth background to the scattering amplitude rather than generating a distinct resonance peak.

In our calculation, when the complex scaling angle is increased up to $\theta=70^\circ$, a pole can indeed be identified slightly below the threshold, with a width characterized by $\Gamma/2 \sim 150~\mathrm{MeV}$. 
The large imaginary part indicates that this pole is far from the physical region and therefore has no clear experimental signature. 
Consequently, no near-threshold resonance is formed in the $I=1$ $DD^{*}$ channel.
This is consistent with the absence of experimental evidence for an isovector partner $T_{cc}'$. 

Overall, the most pronounced impact of the coupled-channel dynamics is the absence of the two higher states present in single-channel scheme. 
A detailed discussion of the pole evolution in both particle and isospin bases is given in Subsec.~\ref{sec:poltr}.
We also considered the $D$-wave contribution in $T_{cc}$ system. 
The \(S\)-wave and \(D\)-wave \(DD^{*}\)-\(D^{*}D^{*}\) interactions are extracted from the experimental data. 
Our analysis shows that the \(D\)-wave contribution, as a high-order correction with a small contribution, can be negligible.

\begin{table}[htbp]
    \centering
    \renewcommand\arraystretch{1.4}
    \setlength{\tabcolsep}{2mm}
    \begin{tabular}{cccccccc}
        \hline\hline
        System & $\Lambda$ (GeV) & BE & $\Gamma$ (keV) & $\sqrt{\langle r^2\rangle}$ (fm) & $P_1$ & $P_2$ & $P_3$ \\
        \hline
        \hline
        $D D^*-D^* D^*$ & 0.8 & 399.5 keV & 75.2 & 4.48 & 69.8\% & 29.2\% & 1.0\% \\
        & 1.0 & 381.1 keV & 80.3 & 4.49 & 70.2\% & 28.5\% & 1.3\% \\
        & 1.2 & 408.4 keV & 88.0 & 4.28 & 67.9\% & 29.7\% & 2.4\% \\
        \hline
        \hline
        $\bar{B}\bar{B}^*-\bar{B}^*\bar{B}^*$ & 0.8 & 45.5 MeV & -- & 0.67 & 35.4\% & 35.6\% & 29.0\% \\
        & 1.0 & 61.7 MeV & -- & 0.60 & 33.0\% & 33.2\% & 33.8\% \\
        & 1.2 & 93.0 MeV & -- & 0.50 & 30.2\% & 30.4\% & 39.4\% \\
        \hline\hline
    \end{tabular}    
    \caption{Properties of the bound states in the $DD^{*}-D^{*}D^{*}$ and $\bar{B}\bar{B}^{*}-\bar{B}^{*}\bar{B}^{*}$ coupled-channel systems for $\Lambda = 0.8, 1.0, 1.2$ GeV. For the charm sector, $P_{1,2,3}$ refer to $D^0D^{*+}$, $D^+D^{*0}$, and $D^{*0}D^{*+}$ channels, respectively. For the bottom sector, they refer to $\bar{B}^-\bar{B}^{*0}$, $\bar{B}^0\bar{B}^{*-}$, and $\bar{B}^{*-}\bar{B}^{*0}$ channels. ``BE'' is the binding energy, and ``$\Gamma$'' is the decay width}
    \label{tab:com_coup_channels}
\end{table}

\subsubsection{$\bar{B}\bar{B}^{*}-\bar{B}^{*}\bar{B}^{*}$}
\begin{figure}[htp]
	\centering
	\begin{tabular}{cc}
			\includegraphics[width=0.48\textwidth]{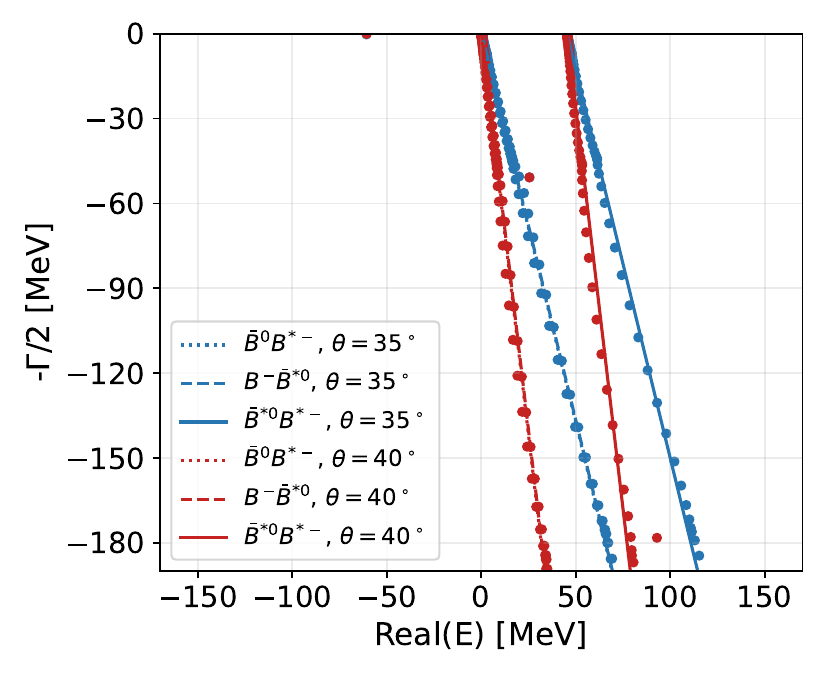}&
			\includegraphics[width=0.48\textwidth]{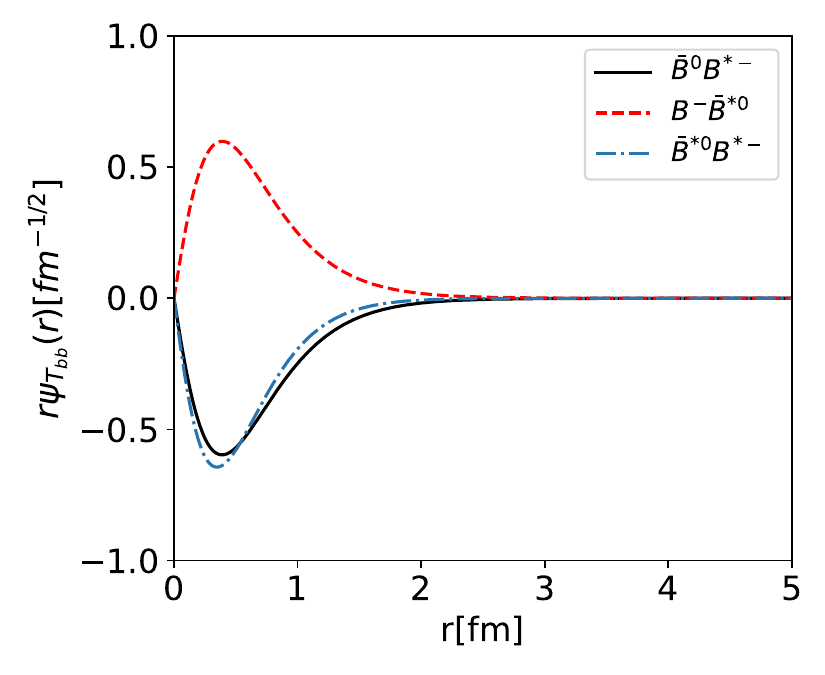}
		\end{tabular}
	\caption{The complex eigenvalues (left) and wave functions (right) of the three components with cutoff $\Lambda=1.0$ GeV for the $\bar B \bar \bar{B}^*$-$\bar \bar{B}^*\bar \bar{B}^*$ system.}
    \label{fig:BBs_BsBs}
\end{figure}

In this subsection, we investigate the $S$-wave $\bar{B}\bar{B}^{*}-\bar{B}^{*}\bar{B}^{*}$ system, which is the bottom analog of the $T_{cc}$ state. 
After considering the coupled-channel effect, we obtain three states in $J^P=1^+$ sector: a deeply bound state and two resonant states.
The corresponding pole positions in the complex energy plane are shown in the left panel of Fig.~\ref{fig:BBs_BsBs}.
The numerical results are summarized in Table~\ref{tab:com_coup_channels} for the bound state.

The bound state is still a almost $I=0$ state.
At $\Lambda=1.0$ GeV, the binding energy is $61.7$ MeV and the root-mean-square radius is $0.6$ fm. 
In the right panel of Fig.~\ref{fig:BBs_BsBs}, we display the wave functions of the channels within the bound state. 
Note that the proportions of the coupling channels \(\bar{B}^{-}\bar{B}^{*0}\), \(\bar{B}^0\bar{B}^{*-}\) and \(\bar{B}^{*-}\bar{B}^{*0}\) are approximately equal, with a ratio of about 1:1:1.
This indicates an important coupled channel effect and the well-preserved isospin symmetry.
Different from its charmed partner, the \(D^*D\) channel is predominant and the contribution from \(D^*D^*\) is comparatively small, only $1.3\%$. 
This is due to the smaller mass gap between the $B$ and $\bar{B}^*$, which is comparable with binding energy.
Besides the bound states, two additional resonances are identified as shown in Fig.~\ref{fig:BBs_BsBs}. 
For a cutoff $\Lambda = 1.0~\mathrm{GeV}$, the lower resonance (Resonance I) has a mass of $10629.5$ MeV and a decay width of $101.6$ MeV almost dominated by the pure $I=1$ component.
Its emergence is associated with repulsive interactions, and the detailed discussion will be presented in Subsec.\ref{sec:poltr}. 
The second resonance (Resonance II)  an \(I=0\) state, related to the single-\(\bar{B}^*\bar{B}^*\) channel, which has evolved from the original $\bar{B}^*\bar{B}^*$ bound state without considering the $B\bar{B}^*-\bar{B}^*\bar{B}^*$ coupled-channel effect in Scheme I.
The coupled-channel effect between \(\bar{B}\bar{B}^*\) and \(\bar{B}^*\bar{B}^*\) channel introduces an effective repulsive interaction, transforming the bound state into a resonance with a significant decay width, primarily decaying into the \(\bar{B}\bar{B}^*\) channel.
Because this resonance has a large decay width, it is difficult to observe and can thus be ignored. 

\subsubsection{$D\bar{B}^{*}-\bar{B}D^{*}-D^{*}\bar{B}^{*}$}
We extend the coupled-channel analysis to the $T_{bc}$ system by considering the interactions among the channels $D^{0}\bar{B}^{*0}$, $D^{+}\bar{B}^{*-}$, $\bar{B}^{0}D^{*0}$, $\bar{B}^{-}D^{*+}$, $D^{*0}\bar{B}^{*0}$ and $D^{*+}\bar{B}^{*-}$.  
Unlike the single-channel case where only $t$-channel dynamics are present, the coupled-channel system naturally allows for $u$-channel contributions. 
Based on the effective Lagrangians in Eq. \eqref{lagrangian:p} and Eq. \eqref{lagrangian:v}, the mixed-flavor transitions, such as the $D\bar{B}^* \to \bar{B}D^*$ channel, correspond to $u$-channel meson exchanges.
Consequently, the coupled-channel effect explicitly incorporates the long-range interactions mediated by $u$-channel pion exchange.

\begin{figure}[htp]
	\centering
	\begin{tabular}{c}
			\includegraphics[width=0.48\textwidth]{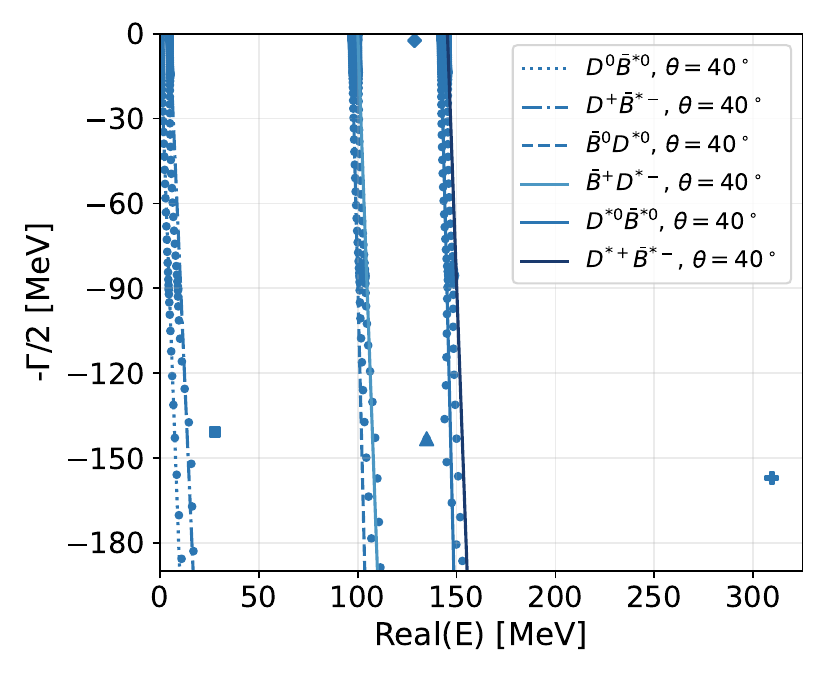}
		\end{tabular}
	\caption{Complex eigenenergies obtained using the CSM for the $D\bar{B}^{*}-\bar{B}D^{*}-D^{*}\bar{B}^{*}$ system with cutoff $\Lambda=1.0$ GeV.}\label{fig:DBs_DsBs}
\end{figure}

For a cutoff  $\Lambda=1.0$ GeV, we obtain a virtual state located $10.96$ MeV below the $D^{0}\bar{B}^{*0}$ threshold, together with four resonances above  the threshold, as illustrated in Figure~\ref{fig:DBs_DsBs}. 
The four resonances arise from the interplay of constructive and destructive interference across the six coupled channels. 
Their properties are summarized in Table~\ref{tab:DBs_DsBs_r}.
The three higher resonances exhibit large decay widths due to strong couplings to multiple channels, making them difficult to observe experimentally. 
The lowest resonance I stands out as an exception.
It lies $31.79$ MeV  below the \(D^{*0}\bar{B}^{*0}\) threshold and has a relatively narrow decay width of $4.8$ MeV. 
Notably, when using Scheme II parameters calibrated to account for inter-channel correlations, we only detected the \(I=0\) virtual state with no evidence of bound states. 
In Ref.~\cite{Liu:2025fhl}, no bound state is found in the coupled-channel system, consistent with our calculation.
\begin{table}
	\centering
	  \renewcommand\arraystretch{1.5} 
        \setlength{\tabcolsep}{3.8mm}  
		\begin{tabular}{ccccc}
         \hline\hline
		          & Resonance I & Resonance II & Resonance III & Resonance IV \\ \hline \hline 
        Mass (MeV) &  $7318.3$   &   7217.3     &     7324.5    &    7499.0    \\ 
		  Width(MeV)  &  $4.8$      &   281.6      &     286.5     &    314.3     \\ 

        \hline\hline
		\end{tabular}
	\caption{Properties of the resonance poles in the $D\bar{B}^{*}-\bar{B}D^{*}-D^{*}\bar{B}^{*}$ system with cutoff values $\Lambda=1.0$ GeV.}
\label{tab:DBs_DsBs_r}
\end{table}

\subsection{Pole trajectories }\label{sec:poltr}
\begin{figure}[htp]
\centering
\begin{subfigure}[t]{0.48\textwidth}
    \centering
    \includegraphics[width=\linewidth]{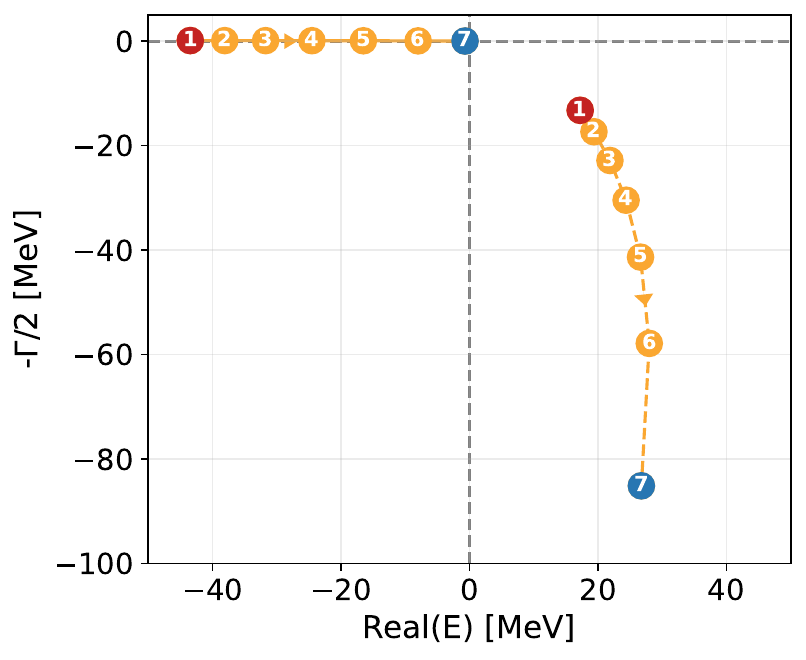}
    \caption{Single-channel system.}
\end{subfigure}
\hfill
\begin{subfigure}[t]{0.48\textwidth}
    \centering
    \includegraphics[width=\linewidth]{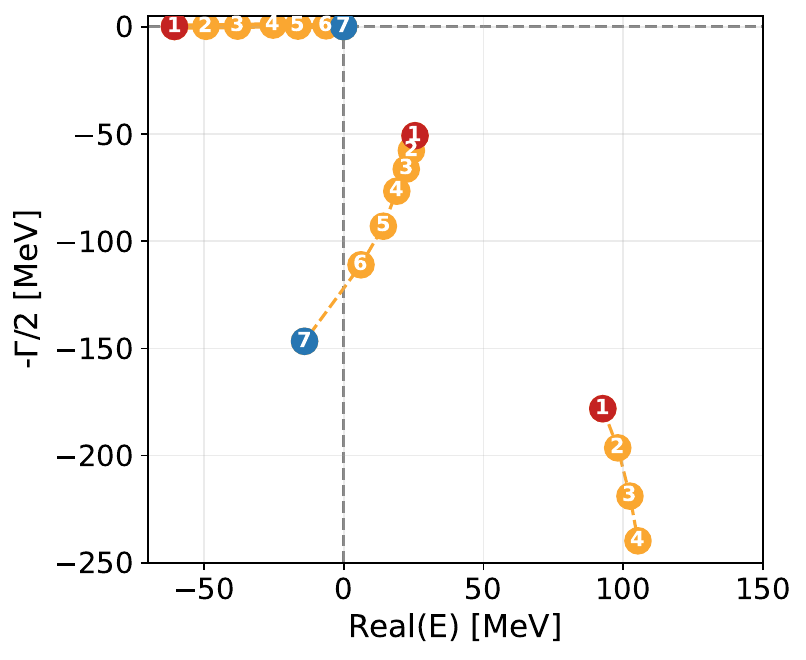}
    \caption{Coupled-channel system.}
\end{subfigure}
\caption{Pole trajectories with varying meson masses. The left panel shows the results in Scheme~I~\cite{Ren:2023pip}, while the right panel shows those in Scheme~II. The circled numbers $1$ to $7$ denote the sequential tuning of the component meson masses from $B^{(*)}$ to $D^{(*)}$.}
\label{fig:poltrl_1}
\end{figure}

In Schemes I, where only $PP^{*}$ interactions are retained, increasing meson masses lead to two different behaviors. 
The isoscalar-dominated bound state exhibits deeper binding, while the first isovector-dominated resonance moves toward the threshold, as shown in Fig.~\ref{fig:poltrl_1} (a). 
Specifically, resonance approaches the threshold with a narrower decay width as the mass parameter increases.
With increasing meson masses, the reduced mass of the system becomes larger, which suppresses the kinetic energy and effectively enhances the role of the interaction. 
As a result, this state eventually transforms into a bound state located on the negative real axis of the complex plane.

After including the $PP^*-P^*P^*$ coupled-channel effect (Schemes II), the evolution of the pole positions from the $D^{(*)}$ to  $\bar{B}^{(*)}$ sector is shown in Fig.~\ref{fig:poltrl_1} (b).
The behavior of the bound state is similar to that in Scheme I, whereas the first resonant state exhibits different behavior.
We find that, as the meson masses increase, this state moves away from the threshold and eventually evolves into a broad subthreshold pole located deep on the unphysical Riemann sheet. 
With a negative real energy and a substantially large imaginary width, it acts only as a smooth background rather than an observable physical state.
This trajectory suggests that the resonance originates from the repulsive potentials~\cite{Hao:2025ukk}.

Compared with the potentials in Table~\ref{tab:obe}, one can see that the main difference between the \(I = 1\) and \(I = 0\) systems lies in the exchange of \(\rho\) and \(\omega\) mesons. 
For the \(I = 0\) system, the \(t -\)channel exchange of vector mesons contributes an attractive interaction. 
Indeed, for the \(I = 1\) state, this exchange leads to a repulsive interaction, and the attractive interaction in the \(u -\)channel related to the coupling constant \(\lambda\) is reduced. 
The coupling constant \(\lambda\) changes from \(0.68\) in Scheme I to \(0.22\) in Scheme II. 
The decay width of the second resonant state is extremely wide and cannot be detected experimentally, so it is not taken into consideration.

\section{Summary}\label{sec:sum}
In this work, we have performed a systematic coupled-channel study of the double-heavy tetraquark $T_{cc}$ and its heavy-flavor counterparts using two fitting schemes. 
Scheme I treats the $DD^*$ system within a single-channel framework, where the coupled-channel effects are implicitly absorbed into short-range interactions. 
Based on HQSS, its parameters are then extended to other single-channel systems, including $D^*D^*$, $\bar{B}\bar{B}^*$, $\bar{B}^*\bar{B}^*$, $D\bar{B}^*$, $D^*\bar{B}$, and $D^*\bar{B}^*$ systems. 
Furthermore, Scheme II explicitly includes the $DD^*$-$D^*D^*$ coupled-channel effects in the fit.
It is observed that the fitted short-range coupling constants are diminished, while the pole position of $T_{cc}^+$ is stable.
%
%
We further extend this coupled-channel framework to the bottom and mixed charm-bottom sectors, comprehensively unraveling the impact of coupled-channel dynamics on the spectrum of doubly heavy tetraquarks.

In the charm ($DD^*$-$D^*D^*$) sector, both schemes yield consistent descriptions of low-lying states, both predicting the $I=0$ $T_{cc}$ as a shallow $DD^*$-dominated bound state, which is actually strongly constrained by the experimental data. 
The significant difference is the fitted coupling constant, which is much smaller in Scheme II. 
The coupled-channel effect also provides the attractive interaction. 
The small coupling constants and additional coupled channel effects change the picture of higher-energy states totally.
Scheme I predicts a $D^*D^*$ bound state due to sufficiently attractive single-channel potentials, while Scheme II shows that $DD^*$-$D^*D^*$ coupling introduces a strong effective repulsion that pushes this bound state completely off the physical sheet, rendering it unobservable. 
For the $I=1$ resonance, it is attractive in Scheme I but becomes effectively repulsive in Scheme II, with its pole moving below threshold and acquiring an extremely large width, becoming an undetectable "sub-pole".

The bottom ($\bar{B}\bar{B}^*$-$\bar{B}^*\bar{B}^*$) sector exhibits superior HQSS and a much smaller inter-channel mass splitting of only about 46 MeV, leading to significantly stronger coupled-channel effects than in the charm sector. 
Nevertheless, low-lying states remain consistent between the two schemes, both predicting a deeply bound $I=0$ state with a binding energy of 40-60 MeV, whose $\bar{B}\bar{B}^*$ and $\bar{B}^*\bar{B}^*$ components are distributed in a near-balanced ratio. 
The most striking feature is the fundamental reversal of the $I=1$ resonance's nature: it is attractive in Scheme I but becomes effectively repulsive in Scheme II, with its formation mechanism completely altered by coupled-channel effects.
Unlike in the charm sector, the overall interaction strength in the bottom sector is sufficiently large that even after becoming repulsive, it still maintains an observable resonance pole.

The mixed charm-bottom ($D\bar{B}^*$-$D^*\bar{B}$-$D^*\bar{B}^*$) sector involves six coupled channels, presenting the most complex dynamics and the most dramatic coupled-channel effects. 
Both schemes agree on the lowest-lying state, predicting a virtual state near threshold. 
However, for higher-energy states, Scheme I predicts a total of five low-lying states, including two additional virtual states each in the $D^*\bar{B}$ and $D^*\bar{B}^*$ channels. 
In sharp contrast, once inter-channel couplings are included in Scheme II, all these extra virtual states acquire decay widths exceeding 200-300 MeV, becoming completely buried in the background and no longer corresponding to identifiable physical states.
Ultimately, only one resonance remains potentially observable.

This study establishes three general patterns governing how coupled-channel dynamics modify the properties of doubly heavy tetraquarks.
\begin{itemize}
    \item Energy dependence: Low-lying near-threshold states are insensitive to coupled-channel structures and can be reliably described by single-channel frameworks, while higher-energy excited states are extremely sensitive, making single-channel predictions potentially completely erroneous. 
    \item General role of coupled channels: They typically introduce effective repulsion that counteracts the internal attraction of single channels, causing single-channel-predicted bound states to become shallower, virtual, resonant, or even disappear entirely. 
    \item Heavy quark mass dependence: Larger heavy quark masses lead to stronger coupled-channel effects but also stronger overall interactions, resulting in deeper low-lying states. 
\end{itemize}
Our results demonstrate that coupled-channel dynamics are indispensable for describing exotic hadrons, and while single-channel frameworks are adequate for the lowest near-threshold states, explicit treatment of coupled-channel effects is essential for obtaining the correct physical picture of all excited doubly heavy tetraquarks.

\begin{acknowledgments}

This work is partly supported  by the National Natural Science Foundation of China (NSFC) under Grants Nos.~12275046, 12547111, 12175239 and 12221005, and by the KAKENHI under Grant Nos. 23K03427 and 24K17055, and by the National Key Research and Development Program of China under Contract No. 2025YFA1613900, and by the Chinese Academy of Sciences under Grant No. YSBR-101.

\end{acknowledgments}

\begin{appendix}

\section{Potentials in $P^{(*)}P^{(*)}$ channel} \label{app:PP}
The scattering process ${P_1^{(*)}(p_1)P_2^{(*)}(p_2)\to P_3^{(*)}(p_3)P_4^{(*)}(p_4)}$ involves the contributions from both the $t$- and $u$-channel meson exchanges, and the explicit forms of effective potentials are given as follows. 

For the scattering process ${PP^{*}\to PP^{*}}$, the potentials read
\begin{eqnarray}
V_{\pi}^{u}&=&\frac{g^2}{f_\pi^2}\frac{(\epsilon_4^{\dagger}\cdot q)(q\cdot\epsilon_2)}{q^2-m_{\pi}^2},\label{eq:v1}\\
V_{\rho/\omega}^{u}&=&-2\lambda^2g_{V}^2\frac{(\epsilon_4^{\dagger}\cdot q)(q\cdot\epsilon_2)-q^2(\epsilon_4^{\dagger}\cdot\epsilon_2)}{q^2-m_{\rho/\omega}^2},\label{eq:v2}\\
V_{\rho/\omega}^{t}&=&\frac{\beta^2g_{V}^2}2\frac{(\epsilon_4^{\dagger}\cdot\epsilon_2)}{q^2-m_{\rho/\omega}^2}.
\end{eqnarray}

For the scattering process ${PP^{*}\to P^{*}P^{*}}$, the potentials are
\begin{eqnarray}
V_{\pi}^{u}&=&\frac{g^2}{f_{\pi}^2}\frac{[\vec{q}\cdot i({\vec{\epsilon}}_{3}^{\dagger}\times\vec{\epsilon}_2)][\vec{\epsilon}_4^{\dagger}\cdot \vec{q}]}{\vec{q}^2-m_{\pi}^2},\\
V_{\pi}^{t}&=&\frac{g^2}{f_{\pi}^2}\frac{[\vec{q}\cdot i(\vec{\epsilon}_4^{\dagger}\times\vec{\epsilon}_2)][\vec{\epsilon}_3^{\dagger}\cdot \vec{q}]}{\vec{q}^2-m_{\pi}^2},\\
V_{\rho/\omega}^{u}&=&-2\lambda^2g_{V}^2\frac{[\vec{q}\cdot i(\vec{\epsilon}_4^{\dagger}\times\vec{\epsilon}_2)][\vec{\epsilon}_3^{\dagger}\cdot \vec{q}]-[\vec{q}\cdot i(\vec{\epsilon}_4^{\dagger}\times\vec{\epsilon}_3^{\dagger})][\vec{q}\cdot\vec{\epsilon}_2]}{\vec{q}^2-m_{\rho/\omega}^2},\\
V_{\rho/\omega}^{t}&=&-2\lambda^2g_{V}^2\frac{[\vec{q}\cdot i(\vec{\epsilon}_3^{\dagger}\times\vec{\epsilon}_2)][\vec{\epsilon}_4^{\dagger}\cdot \vec{q}]-[\vec{q}\cdot i(\vec{\epsilon}_3^{\dagger}\times\vec{\epsilon}_4^{\dagger})][\vec{q}\cdot\vec{\epsilon}_2]}{\vec{q}^2-m_{\rho/\omega}^2}.
\end{eqnarray}

For the scattering process ${P^{*}P^{*}\to PP^{*}}$, the potentials are
\begin{eqnarray}
V_{\pi}^{u}&=&\frac{g^2}{f_{\pi}^2}\frac{[\vec{q}\cdot i(\vec{\epsilon}_4^{\dagger}\times\vec{\epsilon}_1)][\vec{\epsilon}_2\cdot \vec{q}]}{\vec{q}^2-m_{\pi}^2},\\
V_{\pi}^{t}&=&\frac{g^2}{f_{\pi}^2}\frac{[\vec{q}\cdot i(\vec{\epsilon}_4^{\dagger}\times\vec{\epsilon}_2)][\vec{\epsilon}_1\cdot \vec{q}]}{\vec{q}^2-m_{\pi}^2},\\
V_{\rho/\omega}^{u}&=&-2\lambda^2g_{V}^2\frac{[\vec{q}\cdot i(\vec{\epsilon}_4^{\dagger}\times\vec{\epsilon}_2)][\vec{q}\cdot\vec{\epsilon}_1]-[\vec{q}\cdot i(\vec{\epsilon}_1\times\vec{\epsilon}_2)][\vec{\epsilon}_4^{\dagger}\cdot \vec{q}]}{\vec{q}^2-m_{\rho/\omega}^2},\\
V_{\rho/\omega}^{t}&=&-2\lambda^2g_{V}^2\frac{[\vec{q}\cdot i(\vec{\epsilon}_4^{\dagger}\times\vec{\epsilon}_1)][\vec{q}\cdot\vec{\epsilon}_2]-[\vec{q}\cdot i(\vec{\epsilon}_2\times\vec{\epsilon}_1)][\vec{\epsilon}_4^{\dagger}\cdot \vec{q}]}{\vec{q}^2-m_{\rho/\omega}^2}.
\end{eqnarray}

For the scattering process ${P^{*}P^{*}\to P^{*}P^{*}}$, the potentials are
\begin{eqnarray}
V_{\pi}^{u}&=&\frac{g^2}{f_\pi^2}\frac{[\vec{q}\cdot i(\vec{\epsilon}_4^{\dagger}\times\vec{\epsilon}_1)][\vec{q}\cdot i(\vec{\epsilon}_3^{\dagger}\times\vec{\epsilon}_2)]}{\vec{q}^2-m_{\pi}^2},\\
V_{\pi}^{t}&=&\frac{g^2}{f_\pi^2}\frac{[\vec{q}\cdot i(\vec{\epsilon}_3^{\dagger}\times\vec{\epsilon}_1)][\vec{q}\cdot i(\vec{\epsilon}_4^{\dagger}\times\vec{\epsilon}_2)]}{\vec{q}^2-m_{\pi}^2},\\
V_{\rho/\omega}^{u}(\lambda)&=&-2\lambda^2g_{V}^2\frac{[\vec{q}\cdot i(\vec{\epsilon}_4^{\dagger}\times\vec{\epsilon}_1)][\vec{q}\cdot i(\vec{\epsilon}_3^{\dagger}\times\vec{\epsilon}_2)]-\vec{q}^2 [i(\vec{\epsilon}_4^{\dagger}\times\vec{\epsilon}_1)][i(\vec{\epsilon}_3^{\dagger}\times\vec{\epsilon}_2)]}{\vec{q}^2-m_{\rho/\omega}^2},\\
V_{\rho/\omega}^{t}(\lambda)&=&-2\lambda^2g_{V}^2\frac{[\vec{q}\cdot i(\vec{\epsilon}_3^{\dagger}\times\vec{\epsilon}_1)][\vec{q}\cdot i(\vec{\epsilon}_4^{\dagger}\times\vec{\epsilon}_2)]-\vec{q}^2 [i(\vec{\epsilon}_3^{\dagger}\times\vec{\epsilon}_1)][i(\vec{\epsilon}_4^{\dagger}\times\vec{\epsilon}_2)]}{\vec{q}^2-m_{\rho/\omega}^2},\\
V_{\rho/\omega}^{u}(\beta)&=&\frac{\beta^2g_{V}^2}2\frac{[ i(\epsilon_4^{\dagger}\cdot\epsilon_1)][i(\epsilon_3^{\dagger}\cdot\epsilon_2)]}{q^2-m_{\rho/\omega}^2},\label{eq:v3}\\
V_{\rho/\omega}^{t}(\beta)&=&\frac{\beta^2g_{V}^2}2\frac{[ i(\epsilon_3^{\dagger}\cdot\epsilon_1)][i(\epsilon_4^{\dagger}\cdot\epsilon_2)]}{q^2-m_{\rho/\omega}^2}.
\end{eqnarray}
Here, $m_{\pi/\rho/\omega}$ are the exchanged meson masses and $f_{\pi}=132$ MeV refers to the $\pi$ meson decay constant. The ${\epsilon_i}$ ($i=1,2,3,4$) represents the polarization vector of the vector meson $P_i^{*}$. 
\section{Polarization vectors} \label{app:Polar}
In the helicity basis, the polarization four-vector of a particle with momentum \(\vec{k}\) reads
\begin{align}
\epsilon^{\mu}(\vec{k},\lambda = \pm 1)&=\frac{1}{\sqrt{2}}
\begin{pmatrix}
0 \\
\mp \cos\theta\cos\phi + i\sin\phi \\
\mp \cos\theta\sin\phi - i\cos\phi \\
\pm \sin\theta
\end{pmatrix}, \\
\epsilon^{\mu}(\vec{k},\lambda = 0)&=\frac{1}{m}
\begin{pmatrix}
k \\
E_{k}\sin\theta\cos\phi \\
E_{k}\sin\theta\sin\phi \\
E_{k}\cos\theta
\end{pmatrix}.
\end{align}
For a particle with momentum \(-\vec{k}\), its polarization four-vector is defined as
\begin{align}
\epsilon^{\mu}(-\vec{k},\lambda = \pm 1) &= \frac{1}{\sqrt{2}}
\begin{pmatrix}
0 \\
\pm \cos\theta\cos\phi + i\sin\phi \\
\pm \cos\theta\sin\phi - i\cos\phi \\
\mp \sin\theta
\end{pmatrix}, \\
\epsilon^{\mu}(-\vec{k},\lambda = 0) &= \frac{1}{m}
\begin{pmatrix}
 -k\\
E_{k}\sin\theta\cos\phi \\
E_{k}\sin\theta\sin\phi \\
E_{k}\cos\theta
\end{pmatrix}.
\end{align}

\section{Cutoff dependence} \label{app:Cutoff}
To evaluate the stability of our numerical results and the sensitivity of the predicted states to the regularization procedure, we investigate the cutoff dependence of the pole positions. We vary the cutoff parameter within a physically motivated range, specifically $\Lambda = 0.8$, $1.0$, and $1.2$ GeV. Tables~\ref{tab:tccsin} and~\ref{tab:tbbsin} summarize the properties of the $D^{(*)}D^{*}$ and $\bar{B}^{(*)}\bar{B}^{*}$ systems across these values.
For the $D^{(*)}D^{*}$ systems, the predicted pole positions and root-mean-square radii exhibit only a mild dependence on the cutoff parameter $\Lambda$. 
%
Compared with the $D$-systems, the binding energies and root-mean-square radii of the $\bar{B}^{(*)}\bar{B}^{*}$ systems exhibit cutoff dependence, generally showing deeper binding at larger $\Lambda$ values. 
%
In all cases, the channel proportions are nearly independent of the variation of $\Lambda$. 

\begin{table}
    \centering
    \renewcommand\arraystretch{1.5} 
    \setlength{\tabcolsep}{2.5mm}  
    \begin{tabular}{ccccccc} 
        \hline\hline
        System & $\Lambda$ (GeV) & Mass (MeV) & BE (keV) & $\sqrt{\langle r^2\rangle}$ & $P(D^0D^{*+})$ & $P(D^{+}D^{*0})$  \\ 
        \hline\hline
        ${D}{D}^{*}$ & $0.8$ & $3487.4$ & $387.7$  & $4.8$ fm & $70.0\%$ & $30.0\%$ \\
        & $1.0$ & $3482.1$ & $393.0$  & $4.7$ fm & $70.0\%$ & $30.0\%$  \\
        & $1.2$ & $3483.5$ & $391.6$  & $4.7$ fm & $70.3\%$ & $29.7\%$  \\
        \hline
        ${D}^{*}{D}^{*}$ & $0.8$ & 4015.3 & 1836 & 2.62 fm   &  \\
        & $1.0$ & 4015.1 & 1976 & 2.47 fm &    \\
        & $1.2$ & 4014.6 & 2467 & 2.18 fm &    \\
        \hline\hline
    \end{tabular}
    \caption{Properties of the bound/resonance states in the ${D}^{*}{D}^{*}$ and ${D}{D}^{*}$ systems with three cutoff values $\Lambda=0.8$, $1.0$, and $1.2$ GeV. ``BE'' denotes the binding energy, ``$\Gamma$'' represents the decay width, ``$P$'' stands for the channel proportion.}
    \label{tab:tccsin}
\end{table}
\begin{table}
	\centering
	  \renewcommand\arraystretch{1.5} 
        \setlength{\tabcolsep}{2.5mm}  
		\begin{tabular}{ccccccccc}
         \hline\hline
		System & $\Lambda$ (GeV)& Mass (MeV) & BE (MeV) & $\sqrt{\langle r^2\rangle}$ & $P(\bar B^{-}\bar B^{*0})$ 
        & $P(\bar B^0\bar B^{*-})$  \\ \hline \hline						
		$\bar{B}\bar{B}^{*}$  &  $0.8$ &  10572.2 & 31.8 & 0.74 fm & $50.3\%$ & $49.7\%$    \\
		& $1.0$ & 10560.1 & 43.9 & 0.61 fm & $50.2\%$ & $49.8\%$   \\
	  & $1.2$ & 10542.0 & 62.0 & 0.51 fm & $50.2\%$ & $49.8\%$  \\ \hline	
		$\bar{B}^{*}\bar{B}^{*}$ & $0.8$ & 10616.8 &  32.6 & 0.73 fm &   &   &    \\
		& $1.0$ & 10604.3 & 45.1 & 0.60 fm &   &     \\
	  & $1.2$ & 10585.5 & 63.9 & 0.50 fm &   &    \\	 	     
        \hline\hline
		\end{tabular}
	\caption{The properties of the bound state in the $\bar{B}^{(*)}\bar{B}^{*}$ systems with three cutoff values 
		$\Lambda=0.8$, $1.0$, and $1.2$ GeV. 
        The script ``BE" denotes the binding energy and ``$P$" stands for the proportion of the channel. }
    \label{tab:tbbsin}
\end{table}

    \begin{table}
    \renewcommand\arraystretch{1.5} 
    \begin{tabular}{p{3.5cm}<{\centering}<{\centering} p{2.2cm}<{\centering}  
                    p{2.5cm}<{\centering} p{2.2cm}<{\centering} p{2.5cm}<{\centering}}
        \hline \hline
        \multirow{2}{*}{$\Lambda$ (GeV)}  & \multicolumn{2}{c}{$DD^{*}$} & \multicolumn{2}{c}{$\bar{B}\bar{B}^{*}$} \\
        \cline{2-3}
        \cline{4-5}
            & Mass(MeV)   & Width(MeV)      & Mass (MeV)        & Width(MeV)  \\ \hline\hline
        0.8 & $3905.4$    & $184.5$      & $10621.3$    & $20.8$ \\
        1.0 & $3902.4$    & $170.3$      & $10621.2$    & $26.4$ \\
        1.2 & $3890.5$    & $128.3$      & $10610.9$    & $17.4$ \\
        \hline \hline
    \end{tabular}
    \caption{Properties of the resonance poles in the $DD^{*}$ and $\bar{B}\bar{B}^{*}$ systems with three cutoff values 
		$\Lambda=0.8$, $1.0$, and $1.2$ GeV.} 
    \label{tab:resTccbb}
\end{table}

The properties of the $I=1$ resonance poles in the $DD^*$ and $\bar{B}\bar{B}^*$ systems at different cutoffs are summarized in Table~\ref{tab:resTccbb}. In the charm sector, the $DD^*$ resonance is characterized by a substantial width.
Such large decay widths are consistent with our previous results. 
In contrast, the $\bar{B}\bar{B}^*$ resonance is significantly narrower. 
Its mass shows stability, particularly within the $\Lambda=0.8$ to $1.0$ GeV range.

\end{appendix}

\bibliography{ref.bib}
\end{document}